\documentclass[twocolumn,astrosymb,trackchanges]{aastex702}
\usepackage{multirow}
\usepackage{amsmath}

\shorttitle{CVs in VLASS, RACS, and VAST}
\shortauthors{Rao et al.}

\begin{document}

\title{Systematic Search for Cataclysmic Variables in Radio Surveys: First Radio Detection of IX Velorum}

\author[0000-0002-1349-8173]{Suhasini S.~Rao}
\affiliation{University of Alberta,
CCIS 4-181, Edmonton, Alberta, Canada}
\email[show]{suhasin1@ualberta.ca}  

\author[0000-0001-6682-916X]{Gregory R.~Sivakoff}
\affiliation{University of Alberta,
CCIS 4-181, Edmonton, Alberta, Canada}
\email{sivakoff@ualberta.ca}

\author[0000-0003-3944-6109]{Craig O.~Heinke}
\affiliation{University of Alberta,
CCIS 4-181, Edmonton, Alberta, Canada}
\email{heinke@ualberta.ca}

\author[0000-0002-8935-9882]{Akash Anumarlapudi}
\affiliation{Department of Physics and Astronomy, University of North Carolina at Chapel Hill, 120 E. Cameron Avenue, Chapel Hill, NC 27599, USA}
\email{akasha@unc.edu}

\author[0000-0002-4405-3273]{Laura Driessen}
\affiliation{Sydney Institute for Astronomy, School of Physics A28, University of Sydney, NSW 2006, Australia}
\email{laura.driessen@sydney.edu.au}

\author[0000-0001-6295-2881]{David L.~Kaplan}
\affiliation{Center for Gravitation, Cosmology, and Astrophysics,
Department of Physics \& Astronomy, University of Wisconsin-Milwaukee, Milwaukee, WI, USA}
\email{kaplan@uwm.edu}

\author[0000-0002-6896-1655]{Patrick A.~Woudt}
\affiliation{Department of Astronomy, University of Cape Town, Private Bag X3, Rondebosch 7701, South Africa}
\email{patrick.woudt@uct.ac.za}
 
\begin{abstract}
We present a systematic search for radio emission from cataclysmic variables (CVs) across four wide-field radio surveys in the $0.7\text{--}4\rm\ GHz$ frequency range: VLASS; RACS; VAST; and EMU. We generated a reliable catalog of 2225 optically-identified CV candidates using a multi-stage cross-match, combined with systematic and manual verification; 1374 are matched to Gaia sources within the radio surveys' footprints. Among these, we detected six CVs in at least one epoch (at $10^{16.5}\text{--}10^{17.5} \rm\ erg\, s^{-1}\, Hz^{-1}$, consistent with the typical population of radio-detected CVs); five had been detected in radio before. IX Vel, the closest nova-like CV, is detected in radio for the first time and shows quasi-persistent low-level emission across a $10\rm\ h$ EMU ($943.5 \rm \ MHz$) observation. Given most epochs detected no CV, we stacked images without radio detections to better constrain time-averaged brightnesses. We created co-added images and placed upper limits on low-level radio emission from each of the 423 individual CVs within the VAST footprint. To constrain the median brightness for the entire CV population, we combined all Gaia-matched CVs in each survey into single population-wide stacked images, detecting no statistically significant emission. Finally, we created population-wide image stacks of the CVs within $100\rm\ pc$, using 14 epochs from 5 CVs in EMU to constrain their median persistent specific luminosity to $<3\times 10^{14} \rm \, erg \, s^{-1} \, Hz^{-1}$.
\end{abstract}


\section{Introduction}\label{sec:introduction} 
Cataclysmic variables (CVs) are binaries comprising of a white dwarf (WD) accreting matter from a low-mass M or K type donor via Roche-lobe overflow \citep[see][for in-depth reviews]{Warner_1995, Hellier_2001}. These binaries have typical orbital periods (P) of $1.3-12 \rm \ h$, featuring a prominent gap in the $\sim2-3 \rm \ h$ range (known as the period gap), and a sharp minimum cutoff at $\rm P\lesssim80 \rm \ min$ \citep{Knigge_2006}. According to the widely accepted standard model for CV evolution \citep[see][for a review]{Knigge_2011}, the angular momentum loss above the period gap is attributed to magnetic braking via magnetic stellar winds from the low-mass donor, whereas gravitational radiation, and supplementary contributions from consequential angular momentum loss \citep{Schreiber_2016}, dominates the angular momentum loss below the period gap. The period gap arises due to the donor star becoming fully convective at $\rm P\approx3\ h$, causing the donor star to shrink, temporarily shutting down mass-transfer via Roche-lobe overflow. The period minimum is ascribed to the donor becoming sufficiently low mass, and thus degenerate, such that further mass loss causes the donor to expand and therefore increase the orbital separation/period.

Based on the magnetic field strength (\textit{B}) of the WD, and hence the accretion state and/or accretion disk geometry, CVs are classified into magnetic CVs (MCVs) and non-magnetic CVs \citep[see][]{Warner_1995}. Non-magnetic CVs have weak magnetic fields of $ B \lesssim 0.1  \rm \ MG$ and have accretion disks that reach the surface of the WD. The matter is transferred onto the surface of the WD through a transition region called the boundary layer. The non-magnetic CVs are further classified into dwarf novae (DNe) and nova-likes (NLs) based on their accretion rates. DNe have relatively lower accretion rates and exhibit periodic outbursts driven by a thermo-viscous disk instability \citep[see][for reviews]{Osaki_1996, Lasota_2001} that transitions the accretion disk between quiescent (low accretion rate) states and outburst (high accretion rate) states. On the other hand, NLs have high accretion rates and stay in a perpetual outburst state \citep{Dubus_2018}. Magnetic CVs have highly magnetic WDs forcing the accretion flow to follow the magnetic field lines \citep{Warner_1995}. These systems are further classified into polars and intermediate polars. Polars are synchronized binaries and have high magnetic fields ($ B\ge 10 \rm \ MG$) that prevent the formation of accretion disks. The accreted matter from the donor is instead directly channeled along the field lines onto the poles of the WD. On the other hand, intermediate polars ($\rm 1 \ MG\le$\, \textit{B}\ $<10 \ $MG) are generally non-synchronous binaries that feature accretion disks that are truncated near the WD, where the magnetic pressure balances the gas pressure and begins channeling accretion.

CVs are generally thought to be WD analogs of low-mass XRBs (LMXBs). Transient radio emission from XRBs is often associated with synchrotron jets \citep[e.g.,][]{GalloFenderPooley_2003, Fender_2004, Gallo_2006}. By association, CVs are expected to produce radio emission as well. However, initial efforts to detect radio emission from CVs were often unsuccessful \citep[e.g.,][]{Cordova_1983, Fuerst_1986, Nelson_Spencer_1988}. A few DNe were detected during outbursts, but their detections were not reproduced, and some studies mentioned a potential $\sim10\%$ source confusion rate \citep[e.g.,][]{Benz_1983, Benz_1996}. However, other WD systems like symbiotic stars, super-soft sources, and novae were observed to launch jets and other outflows \citep[e.g.,][]{Taylor_1986, Tomov_1998, Sokoloski_2008, Chomiuk_2021, McLoughlin_2021}. Therefore CVs were considered unique among accreting compact binaries due to their apparent lack of radio emission, and by extension, synchrotron jets. In fact, \citet{Livio_1999} and \citet{Soker_2004} used the absence of jets in CVs to constrain the physics of jet launching mechanisms, the details of which remains poorly understood.

However, targeted observations of \object{SS Cyg}, a DNe, during one of its optical outbursts in 2008, detected radio emission that provided compelling evidence for transient synchrotron jets \citep{Kording_2008}. Since then SS Cyg has been detected in multiple follow-up observations during outbursts \citep[e.g.,][]{MillerJones_2013, Russell_2016, Mooley_2017, Fender_2019}. Subsequent searches of magnetic and non-magnetic CVs have resulted in the detection of several CVs \citep[e.g.,][]{Coppejans_2015, Coppejans_2016, Barrett_2020, Hewitt_2020, Ridder_2023, Kersten_2025, Ridder_2026}. Although the radio emission from DNe are generally attributed to transient jets, the origin(s) of radio emission from NLs have remained much more unclear, with emission consistent with both coherent and incoherent emission mechanisms \citep[e.g.,][]{Kording_2011, Coppejans_2015}. Radio emission from magnetic CVs has primarily been explained by coherent plasma emission arising from the donor star and/or from the magnetic field interaction between the WD and the donor \citep[e.g.,][]{Barrett_2017, Barrett_2020, Ridder_2026}. Additionally, recent discoveries of some long-period radio transients (LPTs), characterized by coherent radio emissions with periods ranging from minutes to hours \citep[e.g.,][]{HurleyWalker_2024, deRuiter_2025, Horvath_2026}, include systems that have been identified as CVs \citep[e.g.,][]{Anumarlapudi_2025, Rose_2026, Mo_2026, Knigge_2026}.

To date, thousands of CVs have been optically identified \citep[e.g.,][]{Downes_2001, RitterKolb_2003, Guillochon_2017, Inight_2023}, with an estimated local spatial density of CVs of $\sim 5-6 \times 10^{-6} \rm \ pc^{-3}$ \citep{Pala_2020, Canbay_2023}. However, $<50$ CVs have been detected at radio frequencies, $\sim9$ of which are DNe \citep{Kersten_2025}, while the remaining are predominantly MCVs \citep[see][]{Barrett_2020, Ridder_2026, Mo_2026, Knigge_2026}. The disproportionately small number of radio detections, the incomplete understanding of the underlying emission mechanisms, and the recent link to exotic transients like LPTs emphasize the need to expand the sample of radio-detected CVs. Furthermore, while optical, UV, and X-ray observations also provide access to accretion dynamics, radio observations uniquely probe the magnetic field structures of CV, which determine the structure of accretion flows. Therefore, expanding the sample size of radio-detected CVs is essential in understanding the underlying physical mechanisms governing these systems.

In this paper we discuss a systematic search for CVs in four wide-field radio surveys. We detect six CVs in total, including new radio detections of four previously radio-detected CVs, one previous detection of the most distance radio-emitting CV, and the first radio detection of IX Velorum (\object{IX Vel}). The paper is structured as follows: we discuss the creation of a reliable CV catalog in Section~\ref{sec:sampleSelection}; the radio detections in Section~\ref{sec:radioDetection}; image stacking to find low-level emissions in Section~\ref{sec:radioImageStacking}; and the detection rates for future facilities in Section~\ref{sec:detectionRateFuture}. 

\section{Sample Selection} \label{sec:sampleSelection}
\subsection{Parent Catalog, Gaia Cross-match} \label{sec:parentCatGaiaMatch}
We constructed our initial parent catalog of CVs by merging two of the most widely used optical CV catalogs: \cite{Downes_2001} and \cite{RitterKolb_2003} (henceforth, collectively referred to as RKDCat), extracted using \texttt{VizieR} \citep{Ochsenbein_2000}. However, careful visual inspection and systematic cross-verification revealed several significant issues inherent in the RKDCat: (i) duplicate entries with inconsistent data; (ii) sources that have been reclassified as non-CVs in subsequent works; (iii) large positional errors generated during the creation of the RKDCat, and/or during data ingestion into \texttt{VizieR}. To rectify these discrepancies and establish a robust baseline, we executed a multi-stage cross-match and verification pipeline with the SIMBAD catalog \citep{SIMBAD}. While the SIMBAD catalog, a compilation of tables from various surveys and literature, might not be the most accurate for every source, this rigorous and systematic cross-matching provided a baseline for coordinate verification and rejection of misclassified sources. This resulted in a catalog of 2225 CV candidates with substantially improved reliable in both source classification and sky positions. However, the catalog still retains sources that are also classified as other variables; achieving a purer sample would require manual verification against literature for all 2000+ individual sources, which is beyond the scope of this work. A detailed description of the cross-matching pipeline, a decision flowchart used for the construction of our parent catalog, and a sample of the catalog (Table~\ref{tab:RKDCat_MRT}) is given in Appendix~\ref{app:catalogCreation}. In the table, PARindex is a unique identifier given to each source. The full table in its entirety is provided online in a machine-readable format. 

Before cross-matching our parent CV catalog with the Gaia Data Release 3 (DR3) catalog \citep{Gaia_2023}, we excluded CVs associated with globular clusters (GCs), the Large Magellanic Cloud (LMC), or the Small Magellanic Cloud (SMC), as Gaia's distance estimates are unreliable/unavailable at large ($>2\rm\ kpc$) distances. We removed all CVs within the half-light radii of all the known GCs listed in \citet{Harris_1996}, and also rejected sources within the semi-major axis of the LMC and SMC \citep[$\sim162\arcmin$ and $\sim79\arcmin$ respectively; ][]{Cook_2014}. This reduced the catalog to 2064 CVs. We then cross-matched the remaining sources with the Gaia DR3 catalog using an initially generous matching radius of $10\arcsec$ to account for proper motions. Assuming the coordinates of most sources in SIMBAD or the RKDCat are near epoch J2000, we propagated all matched Gaia sources back to epoch J2000 using their measured proper motion, and selected the nearest source as the potential Gaia counterpart. This yielded a sample of 1946 Gaia cross-matched CVs. The reliability of this sample is given in Section~\ref{subsec:gaiaReliability}. Even if the RKDCat- or SIMBAD- reported positions were given at their original observational epoch rather than the standard J2000, a  typical proper motion of $0.1\arcsec \rm \ yr^{-1}$ \citep{Lepine_2002} from a date as far back as the 1970s will only result in a $\leq 3\arcsec$ positional difference. Such a positional offset would not invalidate our cross-match results, given our initial cross-matching radius of $10\arcsec$.

As a significant fraction of CVs lie below the completeness limit of Gaia ($\rm G \approx 21 \ mag$), we removed sources fainter than this limit. We adopted the Bayesian distance estimates from \citet{BailerJones_2021} derived from the geometric prior distribution model from Gaia's astrometric data. Since the distance posteriors for sources with parallax signal-to-noise (SNR) $<1$ (including negative parallaxes seen in Gaia due to zero-point uncertainties) are mostly dominated by priors, we excluded these sources from our sample as well\footnote{Although we excluded sources with unreliable Gaia astrometry from our sample, we tested whether any of the rejected sources have radio counterparts (as described in Section~\ref{sec:radioDetection}). None were detected}. Finally, we only retained sources with $\rm G_{BP}-G_{RP}$ color or $\rm G$-band magnitude estimates. This resulted in a sample of 1468 sources in our catalog.

Since Gaia's inferred stellar colors and luminosities can be affected by dust extinction, we employed a dust extinction correction method outlined by \citet{Rao_2026}. Briefly, we applied extinction corrections using the Bayestar 3D dust map \citep{Green_2019} for sources with $\delta>-30^\circ$ and DECaPS 3D dust maps \citep{Zucker_2025} for sources with $\delta\leq-30^\circ$ and with Galactic latitude $|b|\leq10^\circ$ using each source's sky position and distance, converted the color excess to $\rm A_V$, and used the precomputed \texttt{dustapprox} models \citep{Fouesneau_dustapprox_2022} to correct for the extinction in $\rm G$ and $\rm G_{BP}-G_{RP}$. For the 164 sources that fall outside the coverage of both Bayestar and DECaPS, we do not apply any extinction correction. Since these sources lie away from the Galactic plane, we assume the dust extinction is negligible. This assumption is supported by the off-Galactic-plane sources covered by the Bayestar 3D dust map, which exhibit typical extinction values of $\rm A_V \lesssim 0.3 \rm \ mag$.

\subsection{Gaia Cross-match Reliability} \label{subsec:gaiaReliability}
Because we used the nearest Gaia cross-match, some of these matches could be chance alignments. To calculate the likelihood of a false match and its corresponding cross-match reliability for each source, we estimated the background Gaia source density~($\rho$) using all Gaia sources within a large radius of $1\arcmin$. We then computed the expected number of false matches $ N_{\rm false}=\rho \pi \rm \theta^2$, where $\theta$ is the angular separation between the parent catalog coordinates and the Gaia source position (proper-motion corrected). Since we found that in all cases $N_{\rm false} \ll1$, we computed reliability~($R$) using $R=1-N_{\rm false}$. The calculated reliability per source is a conservative estimate as we accounted for all Gaia sources to estimate $\rho$ (instead of only sources with $\rm G<21\ mag$).

A table containing the parent catalog's sky coordinates, catalog-reported orbital periods, the SIMBAD coordinates, and the Gaia photometric and astrometric data is provided online in its entirety in a machine-readable format. Table~\ref{tab:allCVs_MRT} outlines the structure of this table. The unique identifier column, PARindex, provides matching indices Table~\ref{tab:RKDCat_MRT}. For our final sample, we only use the 1374 sources with $ R\ge95\%$ as sources with reliable Gaia counterparts.

\begin{deluxetable*}{clll}
\tabletypesize{\footnotesize}
\tablecaption{A descriptive summary of the final CV catalog created for this study.
\label{tab:allCVs_MRT}}
\tablehead{
   \colhead{Row number} & \colhead{Units} & \colhead{Label} & \colhead{Description}
}
\startdata
1 & \nodata & PARindex & Source identifier \\
2 & \nodata & CatalogName & Source name in the original catalog \\
3 & \nodata & CatalogAltName & Alternate name in the original catalog \\
4 & deg & CatalogRAdeg & Right ascension in decimal degrees (J2000.0) in the original catalog \\
5 & deg & CatalogDEdeg & Declination in decimal degrees (J2000.0) in the original catalog \\
6 & day & Period & Orbital period in the original catalog \\
7 & \nodata & Catalog & Original catalog \\
8 & \nodata & SimbadName & SIMBAD name \\
9 & deg & SimbadRAdeg & SIMBAD right ascension in decimal degrees (J2000.0) \\
10 & deg & SimbadDEdeg & SIMBAD declination in decimal degrees (J2000.0) \\
11 & arcsec & SimbadCatalogAngSep & Angular separation between positions from the original catalog and SIMBAD \\
12 & \nodata & gaiaDesignation & Gaia DR3 designation \\
13 & deg & RAdeg & Gaia DR3 right ascension (J2000) at Ep=2016.0 \\
14 & deg & DEdeg & Gaia DR3 declination (J2000) at Ep=2016.0 \\
15 & arcsec & GaiaAngSep & Angular separation between positions from SIMBAD and Gaia DR3 (proper motion corrected)\\
16 & pc & Dist & Bailer-Jones median geometric distance \\
17 & mas/yr & pmRA & Gaia DR3 proper motion in right ascension \\
18 & mas/yr & e\_pmRA & Gaia DR3 standard error of proper motion in right ascension \\
19 & mas/yr & pmDE & Gaia DR3 proper motion in declination \\
20 & mas/yr & e\_pmDE & Gaia DR3 standard error of proper motion in declination \\
21 & mag & Gmag & Gaia DR3 G-band magnitude (extinction corrected) \\
22 & mag & BP-RP & Gaia DR3 BP-RP colour (extinction corrected) \\
23 & \nodata & GaiaReliability & Reliability of crossmatch between SIMBAD position and proper-motion-corrected Gaia DR3 position \\
\enddata
\tablecomments{This table is published in its entirety in the electronic edition of the {\it Astrophysical Journal}. The form is shown here for guidance.}
\end{deluxetable*}

\section{Radio Detection} \label{sec:radioDetection}
\subsection{Radio Surveys} \label{subsec:radioSurveys}

\begin{figure*}[t]
    \plotone{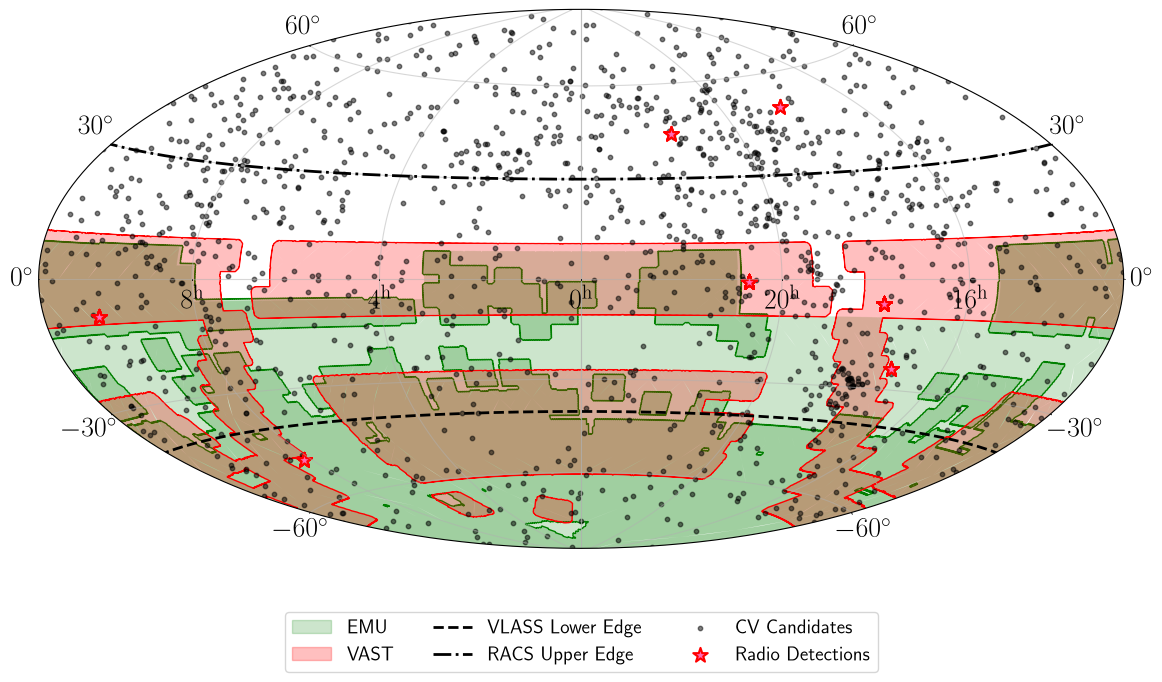}
    \caption{A map of the final sample of CVs. The dot-dashed black curve marks $\delta=-40^{\circ}$, above which VLASS provides full sky coverage. The dashed black curve marks the $\delta=+30^{\circ}$ below which RACS has full sky coverage, good angular resolution and sensitivity. The VAST sky coverage is marked in red; the EMU is shown in green with the current completed regions of EMU shown in dark green ($\approx40\%$ as of July 2026). The final catalog of Gaia matched CVs in shown in black; our sample of radio-detected CVs are shown as red stars.
    \label{fig:skydistribution}}
\end{figure*}

To search for radio counterparts of the CVs, we used wide-field surveys conducted with two telescopes: the upgraded Karl G.~Jansky Very Large Array \citep[VLA;][]{Lacy_2020} and the Australian Square Kilometer Array Pathfinder \citep[ASKAP;][]{Hotan_2021}. Specifically, we used data from the Very Large Array Sky Survey \citep[VLASS; ][]{Lacy_2020}, Rapid ASKAP Continuum Survey \citep[RACS; ][]{McConnell_2020}, ASKAP Variables and Slow Transients survey \citep[VAST;][]{Murphy_2013}, and the Evolutionary Map of the Universe \citep[EMU;][]{Norris_2011}. Together, these surveys provided coverage of the entire sky at different frequencies, sensitivities, and observation cadences. A sky distribution of our final sample of CVs with the sky coverage of each survey is shown in Figure~\ref{fig:skydistribution}. Survey properties are given in Table~\ref{tab:surveyProperties}.

\begin{deluxetable*}{lccccccc}
\tabletypesize{\scriptsize}
\tablecaption{Survey properties of VLASS, RACS, VAST, and EMU.}
\label{tab:surveyProperties}
\tablehead{
\colhead{ } & \colhead{ } & \colhead{ } & \colhead{ } & \colhead{ } & \multicolumn{3}{c}{Positional Uncertainty} \\
\colhead{Survey} & \colhead{Central Frequency} & \colhead{Sensitivity} & \colhead{Sky Area} & \colhead{Beam Size} & \colhead{Statistical} & \colhead{Systematic} & \colhead{Total\tablenotemark{a}} \\
\colhead{ } & \colhead{MHz} & \colhead{$\rm \mu Jy \ bm^{-1}$} & \colhead{deg$^{2}$} & \colhead{arcsec} & \colhead{arcsec} & \colhead{arcsec} & \colhead{arcsec} \\
}
\startdata
VLASS & 3000.0 & 120 & 33,885 & 2.8 & 0.3 & - & 0.5\tablenotemark{b} \\
RACS-low1 & 888.5 & 250 & 33,885 & 18.4\tablenotemark{c} & 1.8 & 0.8 & 1.9 \\
RACS-low2 & 888.5 & 250 & 36,427 & 18.4\tablenotemark{c} & 1.8 & 0.8 & 1.9 \\
RACS-low3 & 943.5 & 250 & 36,427 & 13.4\tablenotemark{c} & 1.3 & 0.8\tablenotemark{e} & 1.6 \\
RACS-mid & 1367.5 & 200 & 36,427 & 10.1\tablenotemark{c} & 1.0 & 2 & 2.2 \\
RACS-high & 1655.5 & 200 & 36,427 & 8.6\tablenotemark{c} & 0.9 & 0.9 & 1.3 \\
VAST & 888.5 & 250 & 12,300 & 15\tablenotemark{d} & 1.5 & 2.5 & 2.9 \\
EMU & 943.5 & 30 & 30,000 & 15\tablenotemark{d} & 1.5 & 0.5 & 1.6 \\
\enddata
\tablenotetext{a}{Total uncertainty is calculated by adding the statistical and systematic uncertainties in quadrature.}
\tablenotetext{b}{VLASS does not provide a systematic uncertainty estimate; however, we use the conservative total uncertainty estimate reported in the VLASS QuickLook User Guide.}
\tablenotetext{c}{As reported by \citet{Duchesne_2025}.}
\tablenotetext{d}{Since the survey is on-going and no empirical median beam size has been reported in literature yet, we used the conservative beam size of $15\arcsec$. }
\tablenotetext{e}{Since systematic uncertainty for RACS-low3 has not been reported yet, we used the uncertainty reported for RACS-low1 and RACS-low2; we take a conservative stance on the uncertainty by not estimating the decreased size due to the small increased frequency of the RACS-low3 observation band.}
\end{deluxetable*}

\subsubsection{VLASS} \label{subsubsec:VLASS}
VLASS was conducted using the upgraded Jansky VLA located in New Mexico, USA. VLASS is a three-epoch\footnote{To address minor issues in the first half-epoch of VAST data, that VLASS epoch was recently observed for a fourth time. However, since this fourth epoch was incomplete at the start of this work, and since the fourth epoch only covers half of the VLASS footprint, we only used the data from the first three fully observed epochs.} S-band ($2-4 \rm \ GHz$) survey that covers the entire sky north of a declination, $\delta$, of $-40^\circ$ ($82\%$ of the sky) at $2\farcs5$ angular resolution. VLASS performs a raster scan using an On-The-Fly (OTF) mosaicking mode, where each pointing gets an effective integration time of $\sim5 \rm \ seconds$. This results in a typical root-mean-square (RMS) sensitivity of $120 \rm \ \mu \rm Jy/bm$. We used the Quick-Look (QL) Stokes~I sky image data products, designed to identify radio transients, for our study. In total, there are 1195 CVs in our final sample that are within the VLASS footprint.

\subsubsection{RACS} \label{subsubsec:RACS}
RACS is a multi-epoch survey conducted using ASKAP, located in Western Australia. It is observed at three spectral bands: RACS-low centered at $887.5 \rm \ MHz$; RACS-mid at $1367.5 \rm \ MHz$; and RACS-high at $1655.5 \rm \ MHz$. We used the two epochs of RACS-low at the original frequency of $887.5 \rm \ MHz$ \citep{Hale_2021, Duchesne_2026}, a third epoch (RACS-low3) at a higher frequency of $943.5 \rm \ MHz$\footnote{RACS-low3 moved to a slightly higher frequency, to obtain improved sensitivity, and to a more compact beam footprint that is more consistent with RACS-mid and RACS-high observations.} (Galvin et al., in prep), two epochs of RACS-mid \citep{Duchesne_2023, Duchesne_2024}, and one epoch of RACS-high \citep{Duchesne_2025} that were complete at the time of this study. Although all RACS epochs except RACS-low1 ($\delta \lesssim +40^\circ$) cover the sky below $\delta \lesssim +50^\circ$ ($\sim88\%$  of the sky), the angular resolution degrades above $\delta >+30^\circ$; we therefore restrict our data selection to below this limit. With an integration time of $\sim15 \rm \ min$ per field, RACS-low epochs achieve a sensitivity of $\sim250 \rm \ \mu Jy/bm$, with RACS-mid and RACS-high epochs reaching $\sim200 \rm \ \mu Jy/bm$. RACS-low, RACS-mid and RACS-high have typical angular resolutions of $\sim18\arcsec$, $\sim12\arcsec$, and $\sim10\arcsec$ respectively. We only use Stokes~I sky images for our study. In total, there are 910 CVs in our final catalog that are within our adopted RACS footprint ($\delta\leq+30^\circ$).

\subsubsection{VAST} \label{subsubsec:VAST}
VAST is an ongoing multi-epoch survey conducted using ASKAP, that aims to observe a portion of the Galactic Plane every two weeks, and select fields of the extragalactic sky every two months, at a $288 \rm \ MHz$ frequency band centered at $887.5 \rm \ MHz$ with a typical angular resolution of $12\text{--}20 \arcsec$ \citep{Murphy_2021}\footnote{\url{https://vast-survey.org/Survey/}}. Each field is observed for $12 \rm \ min$, reaching a typical sensitivity of $\sim250 \rm \ \mu Jy/bm$. We only use the Stokes~I images for our study. In total, there are 423 CVs  in our final catalog that are within the VAST footprint.

\subsubsection{EMU} \label{subsubsec:EMU}
EMU is an on-going wide-field ASKAP survey that observes the entire southern sky ($\delta \le 5^\circ$) in a band $288 \rm \ MHz$ wide centered at $943.5 \rm \ MHz$, with an angular resolution of $15 \arcsec$ \citep{Hopkins_2025}. With a total integration time of $10 \rm \ hours$ per beam, the survey achieves a sensitivity of $\sim30 \rm \ \mu \rm Jy/bm$. At the time of completion of this work (July 2026), $\sim40\%$ of the survey is complete. We use the Stokes~I images, that are split into two 5-hour images, for our study. Although there are 910 CVs from our final sample that are within the full EMU footprint, only 296 are within the fields observed before July 2026, the cut off date adopted for this work.

\subsection{Detection} \label{subsec:radioDetection}
Since source catalogs are only available for some epochs of these radio surveys, we maintain consistency across our work by using forced aperture photometry (FAP) to detect radio emission. In this technique the telescope's point spread function (PSF) is convolved with the image to estimate the flux density of the target point source. We used the proper-motion-corrected Gaia positions to perform the photometry, and detected $\geq5\sigma$ radio emission at the sky position coincidental with 99 CVs in our final catalog in at least one epoch of all four surveys. Of these, we detected 9 in VLASS, 14 in RACS, 47 in VAST, and 26 in EMU. 

\subsection{False Match Rejection and Reliability} \label{subsec:falseMatchRejectionReliability}
Radio detections at the sky positions of our sources do not always guarantee a physical association with our sources, as spatially coincident unrelated background sources can also result in $5\sigma$ detections. This is particularly true for the ASKAP surveys, which have a larger synthesized beam size than VLASS. To ensure that detected high signal-to-noise ratio (SNR) sources are really from our sources and not chance alignments, we first found the centroids of the detected sources and calculated the angular offsets from their respective proper-motion corrected Gaia positions. We calculated the 1$\sigma$ statistical astrometric uncertainties for a $5\sigma$ detection for all surveys/sub-surveys using their median beam size. We then added the survey-specific systematic uncertainty, reported in the literature (see Section~\ref{subsec:radioSurveys}), in quadrature, to calculate the total uncertainty for each survey/sub-survey. The median beam size, systematic uncertainty, statistical uncertainty and total uncertainty for each survey/sub-survey are given in Table~\ref{tab:surveyProperties}. For VLASS, we adopted a more conservative total uncertainty of $0\farcs5$, as reported in the VLASS QuickLook User Guide\footnote{\url{https://science.nrao.edu/science/surveys/vlass/vlass-epoch-1-quick-look-users-guide}}. We used 3 times the total positional uncertainties ($\sigma_{\rm pos}$) as thresholds to retain sources that have a Gaia---radio offset less than the thresholds for each corresponding survey/sub-survey, and thus rejected 1 source from VLASS, 9 from RACS, 27 from VAST and 25 from EMU. Among the retained sources, BB Dor is the only source that had a range of offsets that exceeded the threshold for some epochs in VAST, while being below the threshold for other VAST epochs. A visual inspection of the radio source revealed that the radio emission likely arises from a background AGN; we removed this source from our sample as well. We retained 8 sources in VLASS, 5 in RACS, 19 in VAST, and 4 in EMU.

After visually inspecting these sources, we further removed 2 sources from VLASS, 3 sources from RACS, 16 sources from VAST, and 2 sources from EMU detections. These sources were affected by coincidences with background extended sources like  lobes of radio galaxies and complex filament structures near the Galactic plane, spurious local maxima in the background noise levels, and/or imaging artifacts (primarily linear artifacts in VLASS that arise due to incomplete image deconvolution). After all source rejections, we retained a total of 9 sources that were detected at least once in our surveys, with 6 detected in VLASS, 2 in RACS, 3 in VAST, and 2 in EMU. 

We cross-verified the source type for each of these 9 sources. Among them, \object{EQ CMa} (PAR1325) and \object{V601 Sco} (PAR1981) are also classified as an RS CVn variable \citep[see][]{Malkov_2006} and a Cepheid variable \citep[see][]{Samus_2017} respectively; we therefore removed these two sources from our radio-detected sample. Of the seven remaining, \object{RS Oph} is a symbiotic star and a recurrent nova; since symbiotic stars are not considered as CVs, we removed the source from further analysis. However, because we detect RS Oph across all surveys, we provide a brief analysis and detected radio properties (Table~\ref{tab:RSOph_radioDetections}) in Appendix~\ref{app:RS_Oph}. This yielded a total of six radio-detected CVs. 

To estimate the probability that the detected radio emission originates from unrelated background sources, we first calculated the average source density of each survey. The VLASS and RACS catalogs report source counts of $3.3\times10^{6}$ and $3.1\times10^{6}$ with sky coverages of $\sim33885\rm\ deg^2$ and $\sim36200\rm\ deg^2$ respectively \citep[see][]{Gordon_2021, Duchesne_2024}. EMU is estimated to detect $20\times10^{6}$ sources within $\sim30000\rm\ deg^2$ \citep{Hopkins_2025}. Since the VAST survey has not reported a source count estimate, we used the source density from RACS as both surveys reach similar sensitivity limits. Using the angular offset between the proper-motion-corrected Gaia position and the radio centroid position at the observed epoch, and adopting a $3\sigma$ positional uncertainty as the matching radius for each detected source at each epoch, we estimated the expected number of false matches ($\rm N_{false, radio}$) using the equation in Section~\ref{subsec:gaiaReliability}. Since $\rm N_{false, radio} \ll1$ (ranged between $8\times10^{-8}$ and $0.09$), the resulting reliability is $>99.9\%$ for all radio detections of the six sources (seven, if RS Oph is included as well), suggesting true physical associations. 

\section{Radio Detection Summary} \label{sec:detectionSummary}
Of the six detected CVs, one is a previously radio-detected DN (SS Cyg); two are NLs (\object{RW Sex} and IX Vel), with IX Vel detected for the first time; two are previously radio-detected magnetic CVs (\object{AM Her} and \object{V2400 Oph}); and one is a previously radio-detected propeller system (\object{AE Aqr}). A radio detection summary of these sources including the radio centroid positions, angular separations between the radio centroid and the Gaia position, observation times, flux densities and their respective specific radio luminosities is given in Table~\ref{tab:radioDetections}. Throughout this work, we assume a flat spectral index across the radio frequencies. Although not the most accurate, especially considering the diverse emission mechanisms operating at different frequencies, this assumption allows for a simplified comparison of specific radio luminosities across different surveys and observation frequency bands. We note that except for VLASS detections of AM Her and V2400 Oph --- first reported by \citet{Ridder_2023} --- all remaining detections presented in this work are new. We also note that we do not report the VLASS detection of \object{QS Vir} \citep[a pre-CV first reported by][]{Ridder_2023}. This is because QS Vir was omitted from the primary table of CVs in the \citet{RitterKolb_2003} catalog due to the source nature of QS Vir falling outside the classical definition of a CV. 

\begin{figure*}[t]
    \plotone{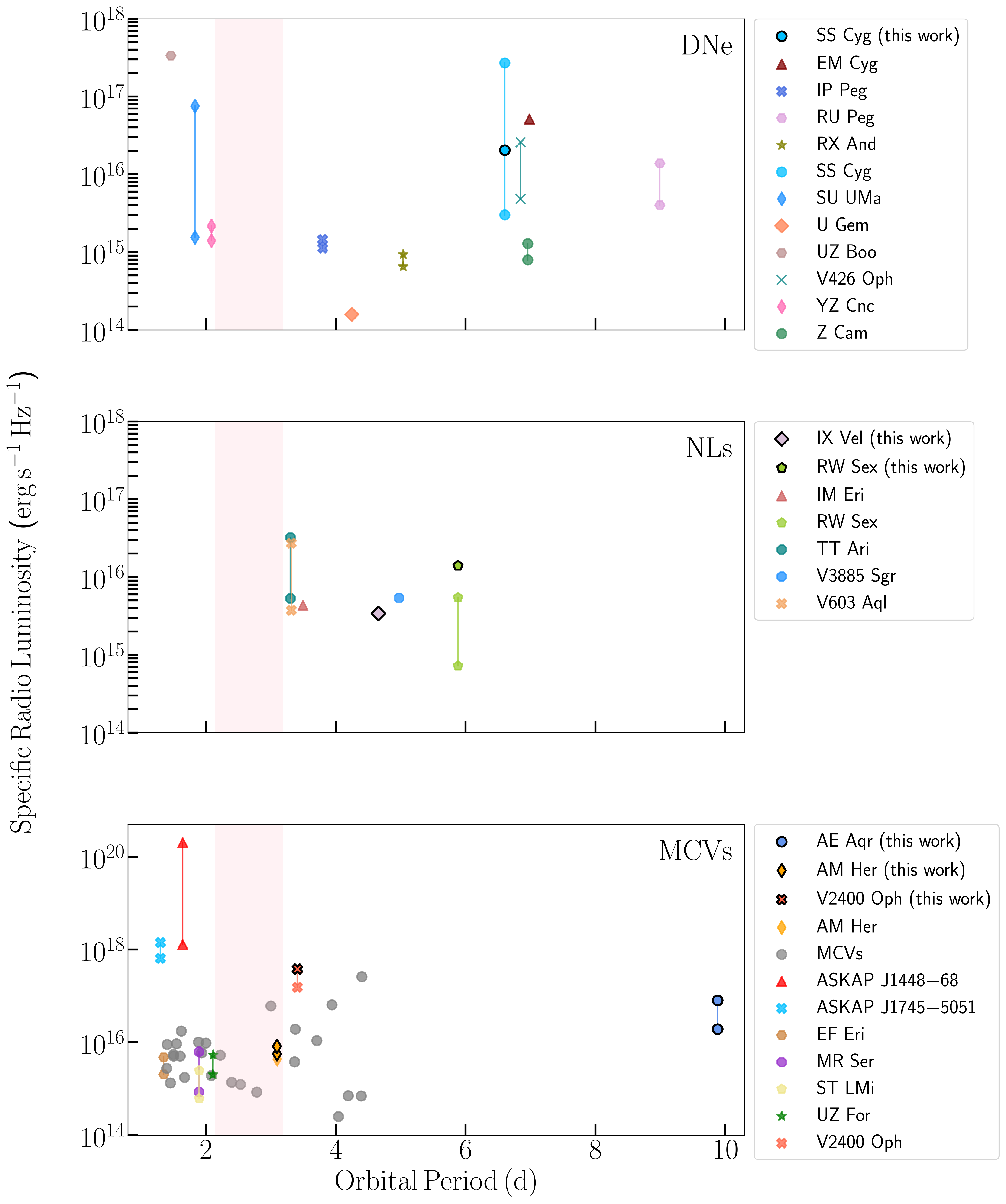}
    \caption{Specific radio luminosity as a function of orbital period for DNe (panel 1), NLs (panel 2), and MCVs (panel 3). Notable CVs of each sub-class are shown in different markers and colors. Radio detections from this work are highlighted with black outlines. Sources with multiple detections are shown as a range instead of marking all detections. The period gap is shown in pink. Note that for ASKAP J1745-5051 we have adopted the poorly constrained distance of $\gtrsim2.7$ kpc \citep{Rose_2026} to estimate its specific luminosity.
    \label{fig:LrPeriod}}
\end{figure*}

A comparison of the specific radio luminosities of our sources with previous reported detections and/or sources, categorized by subtypes, is shown in Figure~\ref{fig:LrPeriod}, with detections from this work highlighted in black. All known radio detections of CVs are included here \citep[see][]{Kording_2008, Kording_2011, Coppejans_2016, Russell_2016, Mooley_2017, Barrett_2020, Hewitt_2020, Ridder_2023, Kersten_2025, Ridder_2026}. Prominent, well known CVs are individually marked as well. We included AE Aqr within the MCV sub-type to avoid defining a special category for a single object. For sources with multiple detections, a range is provided instead of individual data points. The period gap is marked in pink. As expected, and seen before \citep[e.g.,][]{Ridder_2023, Kersten_2025}, we do not see any correlation between the luminosity and orbital period across any sub-types. All radio detections from this work fall within the typical range of radio luminosities previously observed for CVs.  

\startlongtable
\begin{deluxetable*}{llllcccccc}
\tabletypesize{\footnotesize}
\tablecaption{Radio properties of the multi-epoch detections of CVs.
\label{tab:radioDetections}}
\tablehead{
\colhead{PARindex} & \colhead{Name} & \colhead{CV type} & \colhead{Survey} & \colhead{Obstime} & \colhead{Radio R.A.} & \colhead{Radio Dec} & \colhead{Sep\tablenotemark{a}} & \colhead{$\rm S_{\nu}$} & \colhead{$\rm L_{R, \nu}$} \\
\colhead{ } & \colhead{ } & \colhead{ } & \colhead{ } & \colhead{ } & \colhead{deg} & \colhead{deg} & \colhead{arcsec} & \colhead{$\rm mJy$} & \colhead{$\rm erg \, s^{-1} \, Hz^{-1}$}
}
\startdata
PAR440 & SS Cyg & DN & VLASS & 2023-02-13T16:53:08.700 & 325.67933 & 43.58629 & 0.05 & 1.35 $\pm$ 0.13 & 2.0 $\pm$ 0.2 $ \times 10^{16}$ \\
\hline
\multirow{2}{*}{PAR140} & \multirow{2}{*}{AM Her} & \multirow{2}{*}{MCV} & \multirow{2}{*}{VLASS} & 2017-09-09T05:15:42.525 & 274.05506 & 49.86821 & 0.52 & 0.62 $\pm$ 0.12 & 6 $\pm$ 1 $ \times 10^{15}$ \\
& & & & 2020-08-11T05:29:52.350 & 274.05492 & 49.86810 & 0.30 & 0.89 $\pm$ 0.14 & 8 $\pm$ 1 $ \times 10^{15}$ \\
\hline
PAR547 & V2400 Oph & MCV & VLASS & 2018-02-05T13:46:41.925 & 258.15163 & -24.24586 & 0.56 & 0.65 $\pm$ 0.12 & 3.8 $\pm$ 0.7 $ \times 10^{17}$ \\
\hline
PAR342 & IX Vel & NL & EMU & 2023-11-09T16:28:12.100 & 123.82874 & -49.22303 & 0.05 & 0.35 $\pm$ 0.05 & 3.4 $\pm$ 0.5 $ \times 10^{15}$ \\
\hline
PAR428 & RW Sex & NL & EMU & 2026-01-30T13:56:02.800 & 154.98589 & -8.69915 & 0.56 & 0.24 $\pm$ 0.04 & 1.4 $\pm$ 0.2 $ \times 10^{16}$ \\
\hline
\multirow{37}{*}{PAR128} & \multirow{37}{*}{AE Aqr} & \multirow{37}{*}{propeller} & VLASS & 2017-12-16T00:27:30.375 & 310.03849 & -0.87077 & 0.12 & 2.34 $\pm$ 0.14 & 2.3 $\pm$ 0.1 $ \times 10^{16}$ \\
& & & RACS & 2019-04-27T21:16:48.885 & 310.03839 & -0.87041 & 1.43 & 4.8 $\pm$ 0.33 & 4.8 $\pm$ 0.3 $ \times 10^{16}$ \\
& & & VAST & 2020-06-19T17:23:46.628 & 310.03849 & -0.87084 & 0.37 & 4.87 $\pm$ 0.26 & 4.9 $\pm$ 0.3 $ \times 10^{16}$ \\
& & & VAST & 2020-06-19T17:51:58.686 & 310.03887 & -0.87050 & 1.48 & 4.81 $\pm$ 0.51 & 4.8 $\pm$ 0.5 $ \times 10^{16}$ \\
& & & VLASS & 2020-07-17T07:46:07.050 & 310.03857 & -0.87077 & 0.02 & 3.84 $\pm$ 0.29 & 3.8 $\pm$ 0.3 $ \times 10^{16}$ \\
& & & RACS & 2021-01-17T06:06:36.600 & 310.03842 & -0.87048 & 1.19 & 2.44 $\pm$ 0.18 & 2.4 $\pm$ 0.2 $ \times 10^{16}$ \\
& & & VAST & 2021-07-23T15:49:35.930 & 310.03863 & -0.87088 & 0.41 & 3.24 $\pm$ 0.25 & $3.2^{0.3}_{0.2}$ $ \times 10^{16}$ \\
& & & VAST & 2021-07-23T15:49:35.930 & 310.03863 & -0.87088 & 0.41 & 3.24 $\pm$ 0.25 & $3.2^{0.2}_{0.3}$ $ \times 10^{16}$ \\
& & & VAST & 2021-07-23T15:49:35.930 & 310.03863 & -0.87088 & 0.41 & 3.24 $\pm$ 0.25 & 3.2 $\pm$ 0.2 $ \times 10^{16}$ \\
& & & VAST & 2021-07-23T16:06:31.164 & 310.03809 & -0.87048 & 2.08 & 3.16 $\pm$ 0.47 & 3.2 $\pm$ 0.5 $ \times 10^{16}$ \\
& & & VAST & 2021-07-29T15:47:50.204 & 310.03812 & -0.87036 & 2.26 & 1.93 $\pm$ 0.19 & 1.9 $\pm$ 0.2 $ \times 10^{16}$ \\
& & & VAST & 2021-08-21T13:54:41.956 & 310.03870 & -0.87085 & 0.47 & 2.18 $\pm$ 0.26 & 2.2 $\pm$ 0.3 $ \times 10^{16}$ \\
& & & VAST & 2021-08-21T14:11:37.191 & 310.03804 & -0.87102 & 2.17 & 2.26 $\pm$ 0.44 & 2.3 $\pm$ 0.4 $ \times 10^{16}$ \\
& & & VAST & 2021-09-21T11:26:07.700 & 310.03852 & -0.87068 & 0.41 & 2.67 $\pm$ 0.24 & 2.7 $\pm$ 0.2 $ \times 10^{16}$ \\
& & & RACS & 2022-01-10T06:25:23.800 & 310.03848 & -0.87096 & 0.82 & 4.38 $\pm$ 0.23 & 4.4 $\pm$ 0.2 $ \times 10^{16}$ \\
& & & RACS & 2022-04-06T22:57:01.300 & 310.03832 & -0.87075 & 1.00 & 3.19 $\pm$ 0.23 & 3.2 $\pm$ 0.2 $ \times 10^{16}$ \\
& & & VLASS & 2023-01-21T20:22:53.400 & 310.03860 & -0.87078 & 0.06 & 5.78 $\pm$ 0.18 & 5.8 $\pm$ 0.2 $ \times 10^{16}$ \\
& & & VAST & 2023-07-04T18:43:12.800 & 310.03864 & -0.87120 & 1.58 & 7.86 $\pm$ 0.25 & 7.8 $\pm$ 0.3 $ \times 10^{16}$ \\
& & & VAST & 2023-07-04T18:43:12.800 & 310.03864 & -0.87120 & 1.58 & 7.86 $\pm$ 0.25 & $7.8^{0.3}_{0.2}$ $ \times 10^{16}$ \\
& & & VAST & 2023-07-04T19:24:11.300 & 310.03917 & -0.87100 & 2.14 & 4.91 $\pm$ 0.35 & 4.9 $\pm$ 0.4 $ \times 10^{16}$ \\
& & & VAST & 2023-07-04T19:24:11.300 & 310.03917 & -0.87100 & 2.14 & 4.91 $\pm$ 0.35 & $4.9^{0.3}_{0.4}$ $ \times 10^{16}$ \\
& & & VAST & 2023-08-29T15:44:31.400 & 310.03837 & -0.87103 & 1.36 & 2.1 $\pm$ 0.39 & 2.1 $\pm$ 0.4 $ \times 10^{16}$ \\
& & & VAST & 2023-08-30T13:15:30.700 & 310.03893 & -0.87044 & 1.59 & 3.0 $\pm$ 0.24 & 3.0 $\pm$ 0.2 $ \times 10^{16}$ \\
& & & VAST & 2023-10-27T09:20:15.800 & 310.03862 & -0.87116 & 1.41 & 3.63 $\pm$ 0.25 & 3.6 $\pm$ 0.2 $ \times 10^{16}$ \\
& & & VAST & 2023-10-27T09:20:15.800 & 310.03862 & -0.87116 & 1.41 & 3.63 $\pm$ 0.25 & $3.6^{0.3}_{0.2}$ $ \times 10^{16}$ \\
& & & RACS & 2024-01-03T06:54:14.500 & 310.03862 & -0.87044 & 1.16 & 8.07 $\pm$ 0.21 & 8.1 $\pm$ 0.2 $ \times 10^{16}$ \\
& & & RACS & 2024-01-03T06:54:14.500 & 310.03862 & -0.87044 & 1.16 & 8.07 $\pm$ 0.21 & 8.0 $\pm$ 0.2 $ \times 10^{16}$ \\
& & & VAST & 2024-04-21T21:28:03.200 & 310.03899 & -0.87067 & 1.29 & 2.09 $\pm$ 0.25 & $2.1^{0.2}_{0.3}$ $ \times 10^{16}$ \\
& & & VAST & 2024-04-21T21:28:03.200 & 310.03899 & -0.87067 & 1.29 & 2.09 $\pm$ 0.25 & 2.1 $\pm$ 0.2 $ \times 10^{16}$ \\
& & & VAST & 2024-04-21T21:28:03.200 & 310.03899 & -0.87067 & 1.29 & 2.09 $\pm$ 0.25 & $2.1^{0.3}_{0.2}$ $ \times 10^{16}$ \\
& & & VAST & 2024-06-18T19:31:52.200 & 310.03848 & -0.87085 & 0.69 & 3.74 $\pm$ 0.27 & 3.7 $\pm$ 0.3 $ \times 10^{16}$ \\
& & & VAST & 2024-08-12T16:13:08.600 & 310.03836 & -0.87103 & 1.42 & 2.37 $\pm$ 0.39 & 2.4 $\pm$ 0.4 $ \times 10^{16}$ \\
& & & VAST & 2024-10-14T13:00:01.200 & 310.03837 & -0.87045 & 1.50 & 6.92 $\pm$ 0.4 & 6.9 $\pm$ 0.4 $ \times 10^{16}$ \\
& & & VAST & 2024-10-18T11:49:05.900 & 310.03845 & -0.87088 & 0.84 & 2.03 $\pm$ 0.25 & $2.0^{0.2}_{0.3}$ $ \times 10^{16}$ \\
& & & VAST & 2024-10-18T11:49:05.900 & 310.03845 & -0.87088 & 0.84 & 2.03 $\pm$ 0.25 & 2.0 $\pm$ 0.3 $ \times 10^{16}$ \\
& & & RACS & 2024-12-11T08:06:46.900 & 310.03851 & -0.87061 & 0.76 & 3.52 $\pm$ 0.2 & 3.5 $\pm$ 0.2 $ \times 10^{16}$ \\
& & & VAST & 2024-12-24T04:49:40.500 & 310.03870 & -0.87115 & 1.40 & 2.79 $\pm$ 0.23 & 2.8 $\pm$ 0.2 $ \times 10^{16}$ \\
& & & VAST & 2025-04-22T22:05:12.800 & 310.03856 & -0.87020 & 2.03 & 2.44 $\pm$ 0.5 & 2.4 $\pm$ 0.5 $ \times 10^{16}$ \\
& & & VAST & 2025-05-02T21:40:48.100 & 310.03859 & -0.87064 & 0.50 & 3.32 $\pm$ 0.23 & 3.3 $\pm$ 0.2 $ \times 10^{16}$ \\
& & & VAST & 2025-07-13T16:12:55.400 & 310.03854 & -0.87104 & 1.14 & 3.39 $\pm$ 0.28 & 3.4 $\pm$ 0.3 $ \times 10^{16}$ \\
& & & VAST & 2025-09-08T15:09:15.800 & 310.03868 & -0.87002 & 2.67 & 2.23 $\pm$ 0.46 & 2.2 $\pm$ 0.5 $ \times 10^{16}$ \\
& & & VAST & 2025-09-08T15:09:15.800 & 310.03868 & -0.87002 & 2.67 & 2.23 $\pm$ 0.46 & $2.2^{0.4}_{0.5}$ $ \times 10^{16}$ \\
& & & VAST & 2025-09-11T12:12:43.600 & 310.03842 & -0.87100 & 1.25 & 2.08 $\pm$ 0.27 & 2.1 $\pm$ 0.3 $ \times 10^{16}$ \\
& & & VAST & 2025-11-03T09:41:36.500 & 310.03881 & -0.87094 & 0.84 & 6.11 $\pm$ 0.43 & 6.1 $\pm$ 0.4 $ \times 10^{16}$ \\
& & & VAST & 2025-11-12T09:18:24.800 & 310.03851 & -0.87056 & 0.89 & 2.42 $\pm$ 0.24 & 2.4 $\pm$ 0.2 $ \times 10^{16}$ \\
& & & VAST & 2026-03-25T23:30:48.600 & 310.03904 & -0.87112 & 1.84 & 3.18 $\pm$ 0.44 & 3.2 $\pm$ 0.4 $ \times 10^{16}$ \\
& & & VAST & 2026-03-30T00:02:47.600 & 310.03893 & -0.87093 & 1.10 & 6.37 $\pm$ 0.38 & 6.4 $\pm$ 0.4 $ \times 10^{16}$ \\
\enddata
\tablenotetext{a}{Angular separation between the proper-motion-corrected Gaia position and the radio centroid position at that observation epoch.}
\end{deluxetable*}

\subsection{IX Vel} \label{subsec:ix_vel}
Outside of DNe outbursts, IX Vel is the optically brightest CV (maximum magnitude 10.0, NL subtype) in the night sky, at a distance of $\sim90\rm \ pc$ with an orbital period of $\sim4.6\rm \ h$ \citep{Haug1988}. Although IX Vel was previously observed by ATCA \citep[at 5.5 GHz][]{Kording_2011} and by MeerKAT \citep[at 1.3 GHz][]{Hewitt_2020}, it was not detected in either search, with upper limits of 0.6 mJy (ATCA) and 42 $\mu$Jy (MeerKAT). The source was not detected in any of the RACS or VAST observations either. However, IX Vel was detected by EMU at 0.35$\pm0.05$ mJy, marking the first radio detection of IX Vel. This detection is likely due to the better sensitivity achieved by EMU compared to the other surveys (factor of $\sim5$ better than RACS and VAST) and some previous targeted observations. We note that although \citet{Hewitt_2020} achieved a factor of 2 better sensitivity than EMU, IX Vel was not detected in their $\approx2\rm \ h$ observation. This was somewhat surprising given that \object{V3885 Sgr}, a more distant  ($\approx133 \rm \ pc$) NL with similar orbital periods and Gaia photometric properties, was detected by MeerKAT.

\begin{figure*}[t]
    \plotone{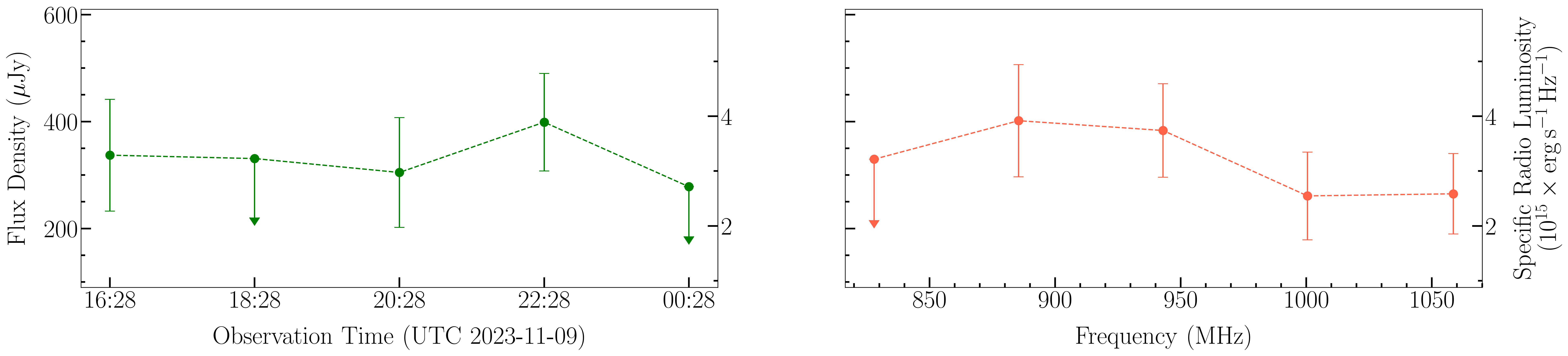}
    \caption{Radio flux density lightcurve (left panel) and spectrum (right panel) of IX Vel during the $10\rm \ h$ EMU observation, with a detection threshold of $3\sigma$. The error bars represent the $1\sigma$ uncertainties in flux density, and upper-limits are shown as downward arrows. The lightcurve shows relatively persistent emission through the observation period; the spectrum is relatively flat ($\alpha\sim-0.5\pm0.5$).
    \label{fig:ixvel_emu}}
\end{figure*}

A dynamic spectra of the EMU observation, generated using \texttt{DSTOOLS} \citep{Pritchard_2025}, showed no significant spectral or temporal features, likely due to the moderate SNR ($\approx 7$) detection of IX Vel during the $10\rm \ h$ observation. We therefore generated a low resolution spectrum and lightcurve (5 sub-intervals each) from the EMU observation, as shown in Figure~\ref{fig:ixvel_emu}. We chose five interval bins to maximize the number of $3\sigma$ detections in each of the sub-images, assuming a constant flux density throughout the observation. We used the calibrated visibility data (obtained from  CSIRO ASKAP Science Data Archive\footnote{\url{https://research.csiro.au/casda/}}) to image the source using \texttt{WSCLEAN} \citep{offringa-wsclean-2014}. The resulting lightcurve and spectrum were extracted in $2\rm\ h$ intervals and $\sim58\rm \ MHz$ frequency bins, respectively, using the FAP method described in Section~\ref{subsec:radioDetection}.

IX Vel exhibits a relatively persistent brightness throughout the observation period. It was not detected in two intervals, but the $3\text{-}\sigma$ upper limits are consistent with our other detections. We did not detect any circularly polarized emission, restricting the circular polarization fraction to $<50\%$. The spectrum does not show any obvious spectral features, and has spectral index of $\alpha=-0.5\pm0.5$ ($\rm S_{\nu} \propto \nu^{\alpha}$). The spectral index is consistent with that of V3885 Sgr \citep[$-0.75 \pm 0.35$ and ${-}0.6 \pm 0.7$ from][respectively]{Kording_2011, Hewitt_2020}. Based on the reported masses of the components of IX Vel \citep[$\rm M_1=0.80^{+0.16}_{-0.11} \ M_{\odot}$ and $\rm M_2=0.52^{+0.10}_{-0.07} \ M_{\odot}$; ][]{Beuermann_1990}, its orbital period, and the time-averaged flux density, we limit the brightness temperature of the emission to $\geq 4\times 10^{10} \rm \ K$. Other than ruling out thermal free-free emission, which generally produces brightness temperatures of $<10^4-10^5\rm \ K$ \citep{Rybicki_Lightman_1979}, we refrain from further constraining the responsible emission mechanism(s) based on the limited data.

\begin{figure*}[t]
    \plotone{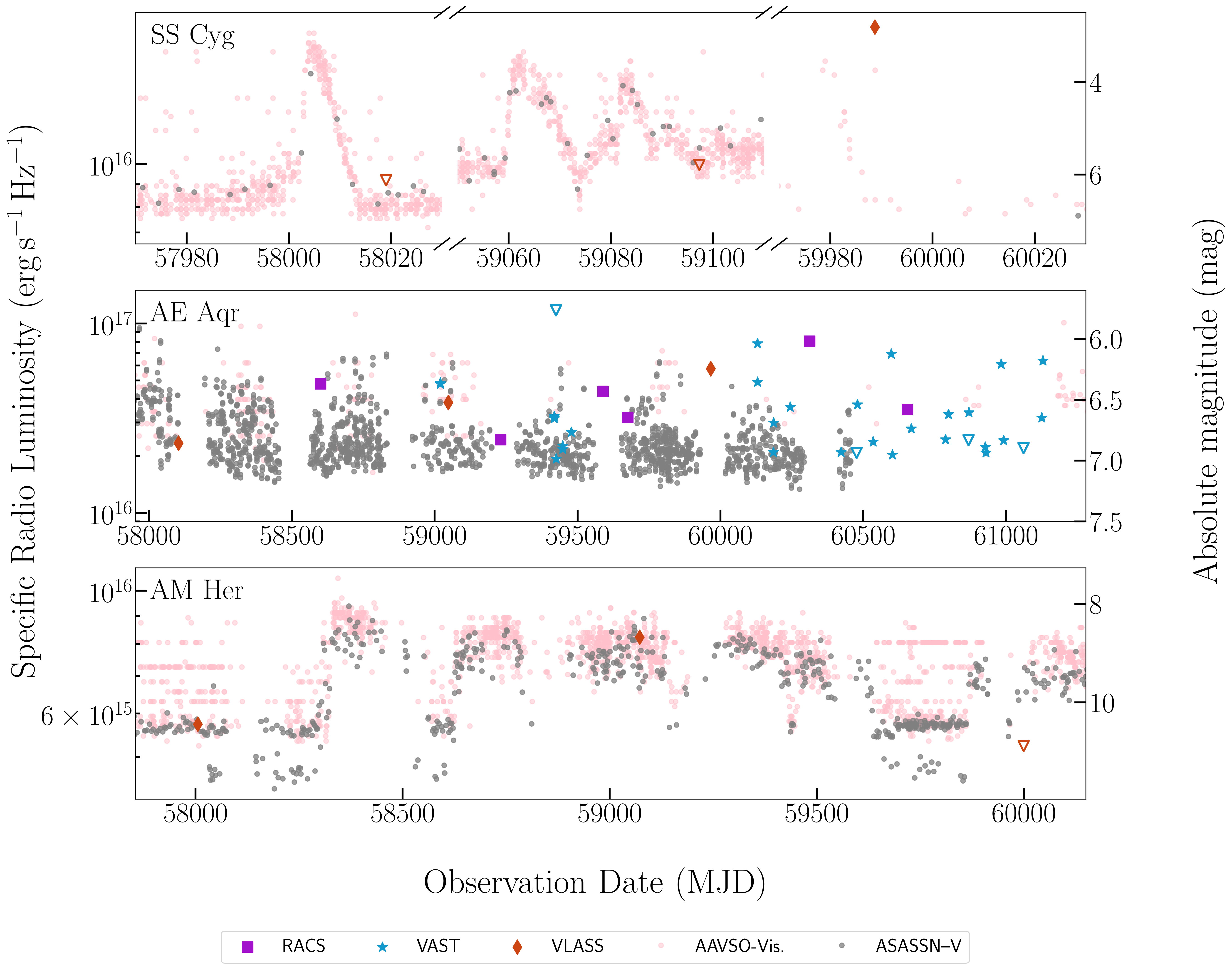}
    \caption{Optical magnitude and specific radio luminosity as a function of time for SS Cyg (panel 1), AE Aqr (panel 2), and AM Her (panel 3). The optical data plotted in the background is AAVSO V-band, shown in pink, and ASAS-SN V-band, shown in grey. Radio upper-limits at epochs of non-detections are shown with inverted triangles. RACS, VAST, and VLASS observations are shown as purple squares, blue stars and red diamonds respectively. The lightcurves for IX Vel, RW Sex and V2400 Oph are omitted due to their singular radio detections, lack of correlation with optical brightness, and/or absence of optical temporal coverage during the radio detections. 
    \label{fig:lightcurve}}
\end{figure*}

\subsection{SS Cyg} \label{subsec:ss_cyg}
SS Cyg is one of the most well studied DNe, and the first radio-detected DN \citep{Kording_2008}. It is at a distance of $\sim 112\rm \ pc$ with an orbital period of $\sim6.6\rm \ h$. Long-term optical observations of SS Cyg have shown a quasi-periodic outburst recurrence timescale of $49\pm15\rm \, d$ and an outburst duration of $\sim7-14\rm \ d$ \citep{Cannizzo_1992}. Since its first radio detections, SS Cyg has been detected multiple times during optical outbursts \citep[see][]{MillerJones_2013, Russell_2016, Mooley_2017, Fender_2019}. We only clearly detected the source in VLASS Epoch 3 ($\sim10\sigma$ detection); the VLASS Epoch 1 observation, which coincided with optical quiescence, resulted in a ($<1\sigma$) non-detection. SS Cyg is also marginally detected at $4\sigma$ in VLASS Epoch 2.

Typically, SS Cyg is considered to be in the decay phase of an outburst when its visual-band (V) magnitude transitions from $\rm V\approx9 \ mag$ to $\rm V\approx11\ mag$ \citep{Cannizzo_1998}, reaching a deep quiescence state at $\rm V\sim11.5\text{--}12 \ mag$. Prior to the Epoch 2 observation, the optical and X-ray quiescent levels following the double outburst (from MJD$\sim$59090, see Figure~\ref{fig:lightcurve}) exhibited significant baseline increase, which prompted an intensive optical follow-up from the AAVSO team \citep{Waagen_2020}. Although this increased accretion flow was speculated to be from X-ray irradiation of the companion star and/or the outer accretion disk, the physical origin of this peculiar behavior remains unconfirmed \citep[see][]{Kimura_2021, Kimura_2023}. Our VLASS Epoch 2 marginal detection is the only reported radio observation of SS Cyg during the shown time range. 

Our single radio detection of SS Cyg (VLASS epoch 3) occurred just after entering the quiescent phase (see Figure~\ref{fig:lightcurve}). This radio detection occurs two hours after a large optical spike ($\rm \Delta V >2 \ mag$) that was reported by a single AAVSO observer. \citet{Mooley_2017} reported a similar late-time radio flaring followed by an optical flare that reached $\approx 4.5 {\rm \, mag}$ absolute magnitude during the February 2016 outburst, approximately 6 days before the source returned to its optical quiescent level. \citet{Fender_2019} also found late-time radio and optical flaring in the April 2016 outburst; however, the flares, with the radio flare being less bright, occurred closer to optical quiescence. Nevertheless, the physical interpretation of the previously observed late-time radio flaring is unexplained. 

On the other hand, if the optical spike is unrelated to the radio detection, the VLASS Epoch 3 detection and VLASS Epoch 2 marginal detection independently provide evidence of radio emission during quiescence. This would place SS Cyg among the very small sample of DNe that are radio-detected during optical quiescence. Currently the sample comprises IP Peg and RU Peg, detected at $4\text{--}5\sigma$ --- although chance alignment cannot be ruled out for RU Peg \citep{Kersten_2025}. Nevertheless, radio emission from DNe during optical quiescence (sometimes even in deep quiescence as seen in IP Peg) hints at non-outburst emission mechanisms operating within these systems that could be responsible for the observed radio emission. 

\subsection{AE Aqr} \label{subsec:ae_aqr}
AE Aqr is a propeller system located at $\sim90\rm \ pc$ with an orbital period of $\sim 10\rm \ h$. Currently there are only two known propellers: AE Aqr and the recently identified \object{LAMOST J024048.51+195226.9} \citep[][]{Pretorius_2021}. Although classified as a CV, thus implying an accreting WD, the WD in AE Aqr accretes little to no material transferred from the donor. Instead, its rapid rotation \citep[33.08 s;][]{Patterson_1979} and high magnetic field ($\sim 1\text{-}10 \rm \ MG$) of the WD results in the expulsion of most \citep[$\sim99.9\%$;][]{Meintjes_2015} of the accreting material from the binary as discrete plasma blobs, through a ``propeller'' effect \citep{Eracleous_1996, Wynn_1997}. These expanding plasma blobs are thought to produce synchrotron emission \citep{vanderLaan_1963, vanderLaan_1966}.

Consistent with this stochastic expulsion of plasma blobs, AE Aqr is observed to be a persistent radio emitter showing rapid variability and flaring behavior \citep{Bookbinder_Lamb_1987, Bastian_1988}. AE Aqr is indeed detected in all observed epochs of VLASS and RACS, as well as most epochs of VAST, displaying varied radio brightnesses (see Figure~\ref{fig:lightcurve}; panel 2). The current footprint of EMU does not cover AE Aqr. A multi-wavelength study of AE Aqr revealed a weak negative correlation between its optical and radio brightness, which was used to suggest a distinction between the optical and radio emission regions \citep{Mauche_2012}. However, because the optical observations are not sampled at the minute-timescale required to sample the flaring, any information about the presence or absence of short timescale optical--radio correlation cannot be evaluated. 

\subsection{AM Her} \label{subsec:am_her}
AM Her is the prototype of the polar sub-class of MCVs. It lies at a distance of $\sim88 \rm \ pc$ and has an orbital period of $\sim3.1 \rm \ h$. We detect AM Her in two VLASS epochs (1 and 2). Even though polars lack accretion disks and do not undergo thermo-viscous outbursts like DNe, they are known to exhibit high- and low-level accretion states, which are reflected in their optical brightness. As seen in Figure~\ref{fig:lightcurve} (panel 3), radio emission from AM Her is seen during both high and low accretion states. This aligns with the accepted interpretation of radio emission from MCVs, where the emission is thought to originate near the donor star due to coronal activity or magnetic interaction between the binary components \citep{Barrett_2017, Ridder_2026} --- processes that are primarily decoupled from the accretion state of the system.

\subsection{RW Sex} \label{subsec:rw_sex}
RW Sex is an optically bright NL at a distance of $\sim 221\rm \ pc$ with an orbital period of $\sim5.8\rm \ h$. It was first detected at radio frequencies by \citet{Coppejans_2015} in two epochs, and was subsequently detected by \citet{Hewitt_2020}. Although the source is within the VLASS and RACS footprints, it was not detected in any of those epochs, likely due to the low sensitivities of those surveys compared to targeted observations. In contrast, RW Sex was detected in EMU at its highest radio luminosity to date ($\rm L_{R,\nu}\approx2\times10^{17} \, erg \, s^{-1} \, Hz^{-1}$), as seen in Figure~\ref{fig:LrPeriod}. Due to the absence of quasi-simultaneous optical coverage during the single EMU detection (in 2026), we cannot compare the radio and optical observations/light-curves.

\subsection{V2400 Oph} \label{subsec:v2400_oph}
V2400 Oph is an intermediate-polar MCV. At a distance of $\sim0.7\rm \ kpc$, it is the farthest CV to be detected at radio frequencies. V2400 Oph has an orbital period of $\sim3.4\rm \ h$. Its VLASS detection was first reported by \citet{Ridder_2023}. It was not detected in the other two epochs of VLASS or any epoch of RACS. Four pointed VLA observations were performed on V2400 Oph, all of which resulted in detections \citep{Ridder_2026}. These detections, in the $100\text{-}700 \rm \ \mu Jy$ range, fall below the $3\sigma$ detection limit of RACS; although EMU would have detected the source at $3\text{-}23\sigma$, the survey does not yet have coverage of this region. Notably, the VLASS detection remains the brightest radio detection of the source to date.

\section{Radio Image-plane Stacking} \label{sec:radioImageStacking}
To probe lower levels of persistent radio emission from the CVs, we combined multi-epoch images to generate a single image (image-plane stacking). Since stacking effectively lowers the RMS noise level (by $\sqrt{\rm N}$, where N is the number of images stacked, in the case of uncorrelated noise), it allows for the detection of any faint but persistent radio emission that would otherwise remain below the detection limit of each individual observation/epoch. For every input image of a source, we extracted $2\arcmin \times 2\arcmin$ cutouts centered around the proper-motion-corrected Gaia positions, so that each pixel across an image stack is at a common offset from the source. We then convolved the images to a common resolution using the \texttt{radio-beam} python package \citep{Koch_2026}. We then calculated the median flux density across images for a given pixel (median stacking) as this technique is less sensitive to outliers (like higher noise levels in individual images) compared to calculating the mean value per pixel. Although, interferometric data can also be added in the visibility space, we chose image-plane stacking since it is significantly less computationally expensive given our sample size. Moreover, \citet{Lindroos_2015} found that both methods yield similar results for point sources; albeit visibility-space stacking produces better SNR compared to image-plane stacking. 

We perform two types of stacking: (1) multi-epoch image stacking within a given survey for individual sources (hereafter producing co-added images); and (2) multi-epoch image stacking of the entire CV population for a given survey (hereafter producing population-wide stacked images).

\subsection{Co-added Images} \label{subsec:VASTimageStacking}
For any individual CV, we can perform image stacking across all observed epochs within a given survey. For our work, we only made co-added images for CVs within the VAST footprint. This is because VAST uniquely provides a high observation cadence within a short period of time (months--years) at a uniform observation frequency band. In contrast, VLASS, RACS and EMU have sparser observations spanning across longer periods, have multiple observation frequency bands, and/or have incomplete survey footprint coverage. 

For a given CV, we co-added input images from only epochs without a ($5\sigma$) detection, in order to probe low-level emission. We also excluded input images with unusually high noise levels --- which we defined as 5 times the median RMS across the images for a given CV --- from the stacking process, as these were typically affected by imaging artifacts that (median) stacking may not fully remove when there are a small number of epochs. To make the co-added images, we used a common angular resolution of $25\arcsec$. For each co-added image of a CV, we applied the bootstrap method to generate a corresponding RMS image for the generated median image. We then applied the FAP method for source detection on the generated median image. Even with co-addition, no additional quiescent emission was detected from any CV above the $5\sigma$ threshold.

\subsection{Population-wide Stacked Images} \label{subsec:avgBrightness}
\begin{deluxetable*}{lccccccccc}
\tabletypesize{\scriptsize}
\tablecaption{Population-wide image stacking properties achieved across different surveys/sub-surveys.}
\label{tab:stackedRMS}
\tablehead{
\colhead{ } & \multicolumn{3}{c}{Full CV Sample} & \multicolumn{5}{c}{CV Sample Within $100 \rm \ pc$} \\
\colhead{Survey} & \colhead{\# CVs} & \colhead{\# Images} & \colhead{Median Image RMS} & \colhead{\# CVs} & \colhead{\# Images} & \colhead{Median Image RMS} & \colhead{Median Dist} & \colhead{$\rm L_{R, \nu}$} \\
\colhead{ } & \colhead{ } & \colhead{ } & \colhead{$\mu\rm Jy/bm$} & \colhead{ } & \colhead{ } & \colhead{$\mu\rm Jy/bm$} & \colhead{pc} & \colhead{$ \times 10^{14} \rm erg \rm \, s^{-1} \rm \, Hz^{-1}$} \\
}
\startdata
VLASS & 1690 & 5066 & 1.6 & 8 & 24 & 21 & 87.1 & $<10$ \\
RACS-low (1 \& 2) & 1368 & 2736 & 3.8 & 4 & 8 & 58 & 87.1 & $<26$ \\
RACS-low3 & 1368 & 1241 & 5.6 & 4 & 4 & 69 & 87.1 & $<31$ \\
RACS-mid (1 \& 2) & 1368 & 2681 & 3.4 & 4 & 8 & 48 & 87.1 & $<22$ \\
RACS-high & 1368 & 1368 & 4.7 & 4 & 4 & 76 & 87.1 & $<34$ \\
VAST & 631 & 18204 & 2.4 & 1 & 21 & 52 & 81.1 & $<20$ \\
EMU & 349 & 740 & 1.6 & 5 & 14 & 8 & 74.7 & $<3$ \\
\enddata
\end{deluxetable*}

In addition to probing individual flux-density limits for CVs, we evaluated the typical radio properties of our entire CV population. Since we center image cutouts on the respective proper-motion-corrected Gaia positions of CVs, we can take the median of input images across multiple sources. We performed such population-wide stacking on all images of all CVs within a survey/sub-survey. To account for the different frequency bands of observations as well as the different sensitivities across the surveys/sub-surveys, we performed the population-wide median stacking strictly for each distinctive survey/sub-survey that were observed at different spectral bands. Within a given survey/sub-survey, we adopted the same restrictions as we did for the co-added images; we only used input images that had not resulted in an individual detection and we excluded input images with an atypically large RMS.

This led to a population-wide images stack of all CVs in: (i) VLASS; (ii) the first two epochs of RACS-low at $888 \rm \ MHz$; (iii) the third epoch of RACS-low at $943.5 \rm \ MHz$; (iv) two epochs of RACS-mid; (v) RACS-high; (vi) VAST; and (vii) EMU images. Except for VLASS --- where all input images were convolved to a common resolution of $5\arcsec$ prior to stacking --- all input images from other surveys were convolved to a common resolution of $25\arcsec$. The chosen common resolution for each survey/sub-survey ensured a uniform beam size across all images/cutouts, while retaining the highest resolution possible. We did not detect any significant emission above the $5\sigma$ detection limit in any population-wide stacked image. The numbers of CV, numbers of input images stacked and the corresponding RMS of the stacked images are given in Table~\ref{tab:stackedRMS}. The number of input images being stacked is typically higher than the number of CVs because (1) VLASS, RACS-low (epochs 1 \& 2), and VAST are designed as multi-epoch surveys; (2) some CVs fall within the overlapping tiles within a given survey; and (3) for EMU, each observation is split into two images, which we treat as two separate epochs.

Since all radio surveys are flux density limited, CVs at larger distances are less likely to be detected. Therefore stacking all images from our entire CV population, without adding a distance limit to account for the flux density limited nature of the radio surveys would bias the stacking towards the non-detection of any low-level persistent radio emission that might be present in these systems. Therefore, to better constrain low-level persistent emission within our sample and get a reasonable estimate of their luminosity, we made population-wide stacked images of all CVs within $100 \rm \ pc$. Again, no average persistent emission was measured above the $5\sigma$ detection limit. The number of images stacked, the RMS of the stacked images, and the $5\sigma$ upper limit (using the median distance of that sample) are also given in Table~\ref{tab:stackedRMS}. Since most of the sources within $100 \rm \ pc$ are within a small range of $70\text{--}100 \rm \ pc$, using the median distances results in minimal bias for the estimation of the average specific radio luminosity upper limits. The tightest constraints, from EMU, suggest that if the local CV population in our $100\rm \ pc$ sample have persistent low-level radio emission, their average specific luminosity must be $\rm L_{R, \nu} <3\times10^{14} \rm \, erg \, s^{-1} Hz^{-1}$. 

\section{CV Detection Rate in Future Radio Surveys} \label{sec:detectionRateFuture}
\begin{figure}[t]
    \plotone{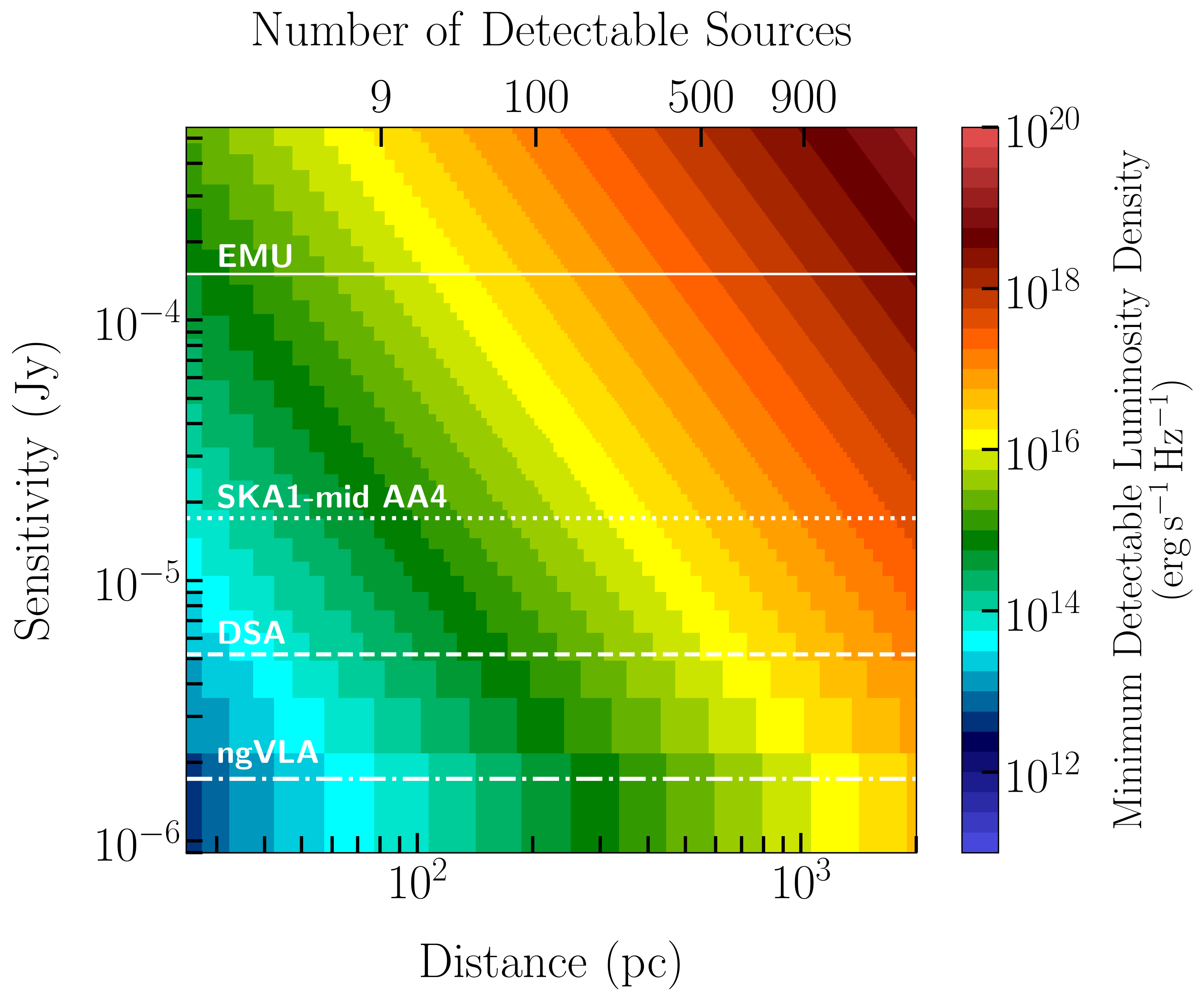}
    \caption{Specific radio luminosities detectable as a function of sensitivity and distance. $5\sigma$ sensitivities of $15\rm \ min$ observations of the next generation of radio telescopes: ngVLA, DSA, and SKA1-Mid AA4 AA4, operating at similar frequencies as current wide-field radio surveys used for this study, are shown with different line markers. 
    \label{fig:future}}
\end{figure}
Figure~\ref{fig:future} shows the different radio luminosities detectable across different sensitivity limits as a function of distance. Among the current generation of radio telescopes and wide-field surveys, EMU offers the best combination of sensitivity ($\approx30\rm\ \mu Jy$) and sky coverage ($\sim75\%$). As shown in the figure, EMU is limited to detecting sources with $\rm L_{R, \nu} \geq 7.5\times10^{15}\ erg \ s^{-1} \, Hz^{-1}$ out to a distance of $100\rm \ pc$. This volume limit restricts our sample to only 21 CVs within our final CV catalog; only 6 of which are within the current EMU footprint. This shows that even the most sensitive current surveys can only probe the uppermost tail of the radio luminosities for CVs, or detect low-level radio emission from only a very small fraction of the nearest sources. On the other hand, targeted surveys conducted using the current radio facilities often achieve higher sensitivities needed to detect more typical low-level radio emission, but have only sampled a small fraction of the CV population, and are frequently designed to be biased towards dwarf nova outbursts, and other CVs known to be radio-bright \citep[e.g.,][]{Barrett_2020, Hewitt_2020, Kersten_2025, Ridder_2026}. 

Next-generation radio facilities/surveys can dramatically transform the detection rates of CVs by bridging the gap between targeted, deeper radio observations and shallow, wide-field surveys. Some upcoming facilities include the Square Kilometer Array Phase 1-Mid \citep[SKA1-Mid AA4; covers $\delta\lesssim+30^\circ$][]{Swart_2022}, the next-generation Very Large Array \citep[ngVLA; covers $\delta \gtrsim-40^\circ$][]{Selina_2018, Murphy_2018}, and the Deep Synoptic Array \citep[DSA; covers $\delta \gtrsim -40^\circ$][]{Hallinan_2019}. These three telescopes, specifically, will operate at similar frequencies as the current wide-field surveys in this paper, but will achieve unprecedented sub-$\rm\mu Jy$ sensitivity levels with angular resolution better than $3\arcsec$. 

The typical sensitivities of SKA1-Mid AA4, DSA, and the ngVLA for $1\rm\ h$ observations is projected to be $\sim2\rm\ \mu Jy$, $\sim0.6\rm\ \mu Jy$, and $\sim0.2\rm\ \mu Jy$ at $1.4\rm \ GHz$, $1.4\rm \ GHz$, and $2.4\rm \ GHz$ respectively. Given the facilities' instantaneous fields-of-view, $15\rm \ min$ exposures over their full visible sky around $\approx2\rm \ GHz$ could be feasible as a large multi-year project with many science goals. In fact, DSA, the instrument with the highest grasp (sensitivity times instantaneous field of view) at $1.5\rm \ GHz$, is already planning its survey consisting of $20\rm\ min$ observing blocks\footnote{The overview of the DSA all-sky survey is available at \url{https://www.deepsynoptic.org/overview}}. We therefore estimate the detection statistics for such surveys. 

A $20 \rm \ min$ integration observation with SKA1-Mid AA4, DSA, and the ngVLA will have sensitivity levels of $\sim3.4\rm\ \mu Jy$, $\sim1\rm\ \mu Jy$, and $\sim0.3\rm\ \mu Jy$ respectively. The $5\sigma$ sensitivities of these surveys are shown as white horizontal lines in Figure~\ref{fig:future}. Within the current completed EMU footprint, we detect two CVs within $225 \rm \ pc$ at $\rm L_{R, \nu} \geq 3\times10^{15}\ erg \ s^{-1} \, Hz^{-1}$. Within this volume, our catalog contains 124 CVs, 39 of which are within the current completed EMU footprint, suggesting that $\sim5\%$ ($\approx 2/39$) of the local CV population are above this luminosity limit. At this specific radio luminosity, sources are detectable out to $\sim384\rm\ pc$, $\sim708\rm\ pc$, and $\sim1293\rm\ pc$ with SKA1-Mid AA4, DSA, and the ngVLA, covering 218, 591, 877 CVs within their respective footprints. Assuming the $\sim4.5\%$ detection fraction across the entire volume, SKA1-Mid AA4, DSA, and the ngVLA will likely detect $\sim11$, $\sim30$, and $\sim44$ CVs respectively at $\rm L_{R, \nu} \geq 10^{15}\ erg \ s^{-1} \, Hz^{-1}$, thus substantially increasing the sample size of radio-detected CVs required for population synthesis studies. 

\section{Conclusion} \label{sec:conclusion}
In this work, we conducted a systematic search for CVs in four wide-field radio surveys: VLASS, RACS, VAST, and EMU. To establish a highly reliable baseline catalog, we developed multi-stage pipelines to construct a robust parent catalog of 2225 CVs from two legacy CV catalogs, which we cross-matched against Gaia to yield a sample of 1374 CVs with Gaia counterparts. Our systematic search for radio counterparts of these CVs resulted in the detection of six CVs: one DN (SS Cyg), two NLs (IX Vel and RW Sex), two MCVs (AM Her, V2400 Oph) and one propeller system (AE Aqr). Of the six CVs, we report new radio detections for four CVs: three previously radio-detected CVs and the first radio detection of IX Vel.

IX Vel shows quasi-persistent radio emission and a roughly flat spectrum across the $10\rm \ h$ observation period of EMU. Due to low SNR, we refrain from speculating origin(s) of the detected radio emission. SS Cyg is detected in VLASS Epoch 3 and marginally detected in VLASS Epoch 2, both immediately after reaching optical quiescence. The VLASS Epoch 3 detection is within 2 hours of a $2 \rm \ mag$ sudden spike in optical emission. Because of the sparse radio and optical observations, we cannot conclude whether the radio emission is analogous to previously observed late-time flarings, or similar to quiescent radio detections of IP Peg and RU Peg. AE Aqr is detected in multiple epochs of VLASS, RACS, and VAST, with its variability consistent with its previous persistent and highly variable radio detections. AM Her is detected in two VLASS epochs during both the high and low accretion states of the system. RW Sex is detected in EMU --- its brightest detection to date. We also confirmed the previous VLASS detection of V2400 Oph, the farthest CV to be detected at radio frequencies (excluding detections during classical nova outbursts, a completely different source of radio emission).

We performed image-plane co-addition of undetected epochs for individual CVs within the VAST footprint, which yielded no statistically significant ($>5\sigma$) detection of persistent low-level radio emission. Similarly, our multi-epoch image stacking of the entire CV population within given survey/sub-survey yielded no detection. Finally, we performed a multi-epoch image stacking of the 21 CVs within the $100\rm\ pc$ volume limited sample; stacking 14 epochs of (5 CVs) EMU images yielded a population-averaged specific luminosity of $3\times10^{14} \rm\  erg \, s^{-1}\, Hz^{-1}$.
 
Future radio facilities like SKA1-Mid AA4, ngVLA and DSA will be able to achieve sub-$\rm\mu Jy$ sensitivities capable of exploring low-level radio emissions from CVs to farther distances, and therefore significantly expanding the current sample of radio-detected CVs.

\begin{acknowledgments}
SSR and GRS are supported by NSERC Discovery Grants RGPIN-2021-04001 and RGPIN-2026-05822.
COH is supported by NSERC Discovery Grant RGPIN-2023-04264, and Alberta Innovates Advance Program 242506334.
SSR, GRS, and COH are members of the University of Alberta. The University of Alberta, its buildings, labs and research stations are primarily located on the territory of N\'ehiyaw (Cree), Niitsitapi (Blackfoot), M\'etis, Nakoda (Stoney), Dene, Haudenosaunee (Iroquois) and Anishinaabe (Ojibway/Saulteaux), lands that are now known as part of Treaties 6, 7 and 8 and homeland of the M\'etis. The University of Alberta and its members respect the sovereignty, lands, histories, languages, knowledge systems and cultures of all First Nations, M\'etis, and Inuit. DK was suported by NSF grant AST-2511757.
This scientific work uses data obtained from Inyarrimanha Ilgari Bundara, the CSIRO Murchison Radio-astronomy Observatory. We acknowledge the Wajarri Yamaji People as the Traditional Owners and native title holders of the Observatory site. CSIRO’s ASKAP radio telescope is part of the Australia Telescope National Facility (https://ror.org/05qajvd42). Operation of ASKAP is funded by the Australian Government with support from the National Collaborative Research Infrastructure Strategy. ASKAP uses the resources of the Pawsey Supercomputing Research Centre. Establishment of ASKAP, Inyarrimanha Ilgari Bundara, the CSIRO Murchison Radio-astronomy Observatory and the Pawsey Supercomputing Research Centre are initiatives of the Australian Government, with support from the Government of Western Australia and the Science and Industry Endowment Fund.
\end{acknowledgments}

\bibliography{mainbib}{}
\bibliographystyle{aasjournalv7.1}

\appendix
\restartappendixnumbering

\twocolumngrid
\section{Catalog Creation}\label{app:catalogCreation}

In the two most widely used optically-selected CV catalog we used for this work, \citet{Downes_2001} and \citet{RitterKolb_2003} (RKDCat), we identified discrepancies across the original catalogs, the \texttt{VizieR} version of the catalogs, and/or the literature used to compile these catalogs. The inconsistencies include:
\begin{itemize}
    \item Rounding/truncation errors on sky positions introduced during the ingestion of the catalogs into \texttt{Vizier}.
    \item Incorrect use of sky coordinate errors/uncertainties as last decimal digit of the coordinate (predominantly declination).
    \item Outdated sky positions.
    \item Typographical errors in sky coordinates and source names.
    \item Duplicate entries with unique source identifiers arising from different studies despite negligible angular separation.
    \item Inclusion of non-CV contaminants; although these sources could have been initially classified as CV candidates at the time of creation of these catalogs.
\end{itemize}

\begin{figure*}[b]
    \plotone{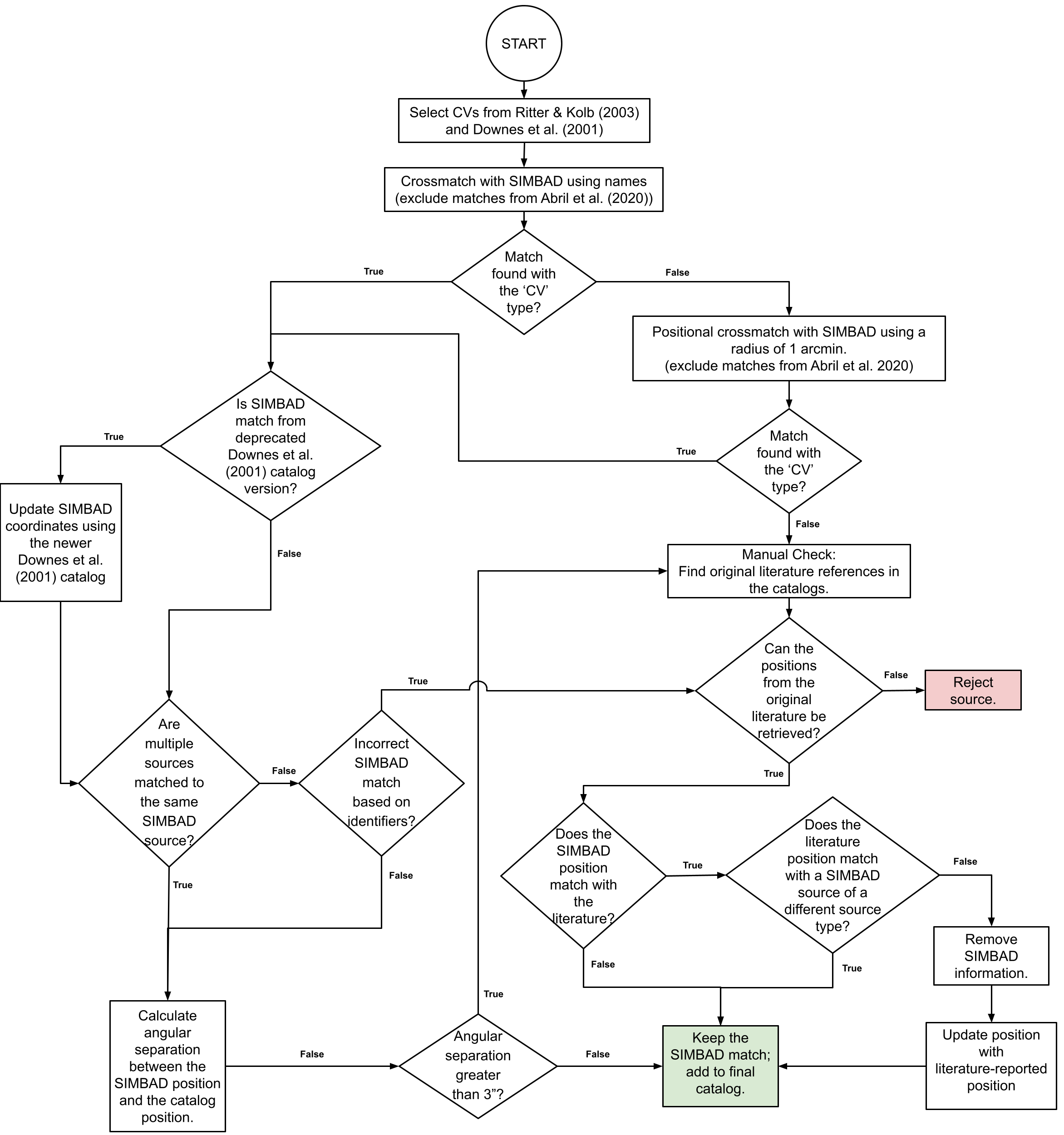}
    \caption{Pipeline used to create the final CV parent catalog.
    \label{fig:CV_flowchart}}
\end{figure*}

Addressing these anomalies was necessary to construct a clean, Gaia cross-matched parent catalog. To rectify these errors to the best of our abilities without going through a time-consuming and rigorous process of manually checking all 2000+ sources, we created the following pipeline. A flowchart of the pipeline is shown in Figure~\ref{fig:CV_flowchart}.

In this pipeline, we either matched each source to a valid SIMBAD source, or, for the sources without a credible SIMBAD-match, we manually verified and corrected the sky position based on the literature reporting the source. The detailed steps of this pipeline is as follows:
\begin{enumerate}
    \item We first attempted cross-matching each source in the RKDCat with the SIMBAD database, using its reported name (or an alternative name, if provided). For sources lacking a SIMBAD match, we subsequently performed a spatial cross-match using their reported sky coordinates and a large matching radius of $1\arcmin$. The use of a relatively large matching radius and the initial cross-matching with name was necessary to avoid spurious cross-matches with unrelated sources arising from the incorrect coordinates reported in RKDCat, as described above. 
    
    \item To ensure reliability, we restricted the SIMBAD cross-matches to only sources classified as a CV (source type ``CV''). Furthermore, SIMBAD sources originating from \citet{Abril_2020} were rejected since they use the \citet{RitterKolb_2003} catalog cross-matched with Gaia without correcting for the above mentioned discrepancies. Therefore, their matched Gaia sources ingested into SIMBAD might not be accurate. Sources without a SIMBAD match after these filters were flagged for manual verification against reporting literature.
    
    \item Since SIMBAD aggregates tables from the literature and surveys, the database retains previous deprecated versions of the RKDCat, primarily the \citet{Downes_2001} catalog. To eliminate outdated sky positions of the SIMBAD-matched sources corresponding to deprecated versions of the \citet{Downes_2001} catalog, we updated the coordinates of all affected sources with their updated coordinates from the latest version of the \citet{Downes_2001} catalog.
    
    \item For multiple sources linked to the same SIMBAD match, we only retained genuine RKDCat-SIMBAD matches verified based on their primary identifiers. RKDCat sources with erroneous SIMBAD matches were stripped of their SIMBAD association and flagged for manual inspection.

    \item To evaluate the cross-match accuracy, we calculated the angular separation between the reported sky coordinates. Given that the earliest literature used in the generation of RKDCat dates to the 1970s, and using the typical proper-motion of $0.1\arcsec \rm \ yr^{-1}$ \citep{Lepine_2002}, we adopted a matching radius of $3\arcsec$ to accommodate proper-motion across 30 years (with respect to epoch J2000.0). Sources with angular separation exceeding this limit were flagged for manual inspection. The retained sources were kept in our final parent CV catalog.
     
    \item For all the sources that were flagged for manual inspection in the above steps, we attempted to retrieve and review the original literature referenced in the RKDCat. We systematically excluded all sources for which the original literature could not be located or verified.

    \item We retained sources for which we could confirm the sky positions from the original literature. 
    
    \item Among the remaining sources that had literature-verified coordinates, we retained sources with a different SIMBAD match with a different source type (predominantly ``eclipsing binary'').

    \item For the final remaining sources, we retained sources with literature-verified coordinates but removed their associated SIMBAD match.
\end{enumerate}
The final RKDCat consists of 2225 CVs. A sample of this catalog is given in Table~\ref{tab:RKDCat_MRT}. The full table in its entirety is provided online in a machine-readable format. 

\begin{longrotatetable}
\movetabledown=20mm
\begin{deluxetable*}{ccccccccccc}
\tabletypesize{\scriptsize}
\tablecaption{RKDCat catalog.
\label{tab:RKDCat_MRT}}
\tablehead{
\colhead{PARindex} & \colhead{CatalogName} & \colhead{CatalogAltName} & \colhead{CatalogRAdeg} & \colhead{CatalogDEdeg} & \colhead{Period} & \colhead{Catalog} & \colhead{SimbadName} & \colhead{SimbadRAdeg} & \colhead{SimbadDEdeg} & \colhead{SimbadCatalogAngSep} \\ 
\colhead{ } & \colhead{ } & \colhead{ } & \colhead{deg} & \colhead{deg} & \colhead{day} & \colhead{ } & \colhead{ } & \colhead{deg} & \colhead{deg} & \colhead{arcsec} 
} 
\startdata
PAR0 & FR Lyn &  & 133.5584269787 & 39.0937085720 & 0.08 & RK/Downes & 1ES 0851+39.2 & 133.5584269787 & 39.0937085720 & 4.4e-01 \\
PAR1 & MV Leo & Leo 5 & 157.0003357834 & 21.8037704335 & 0.15 & RK/Downes & 1H 1025+220 & 157.0003357834 & 21.8037704335 & 5.0e-01 \\
PAR2 & V2731 Oph &  & 262.5912568356 & -5.9921778820 & 0.64 & RK/Downes & 1H 1726-058 & 262.5912568356 & -5.9921778820 & 1.6e-01 \\
PAR3 & J0354-1652 & RBS 490 & 58.5429557606 & -16.8805651156 & 0.03 & RK/Downes & 1RXS J035410.4-165244 & 58.5429557606 & -16.8805651156 & 1.4e-01 \\
PAR4 & V5662 Sgr &  & 301.4630117908 & -29.5833458937 & 0.06 & RK/Downes & 1RXS J200552.6-293455 & 301.4630117908 & -29.5833458937 & 2.3e+00 \\
PAR5 & 0506+7725 &  & 78.4023723178 & 77.4785507301 & 0.15 & RK/Downes & 2E  1224 & 78.4023723178 & 77.4785507301 & 1.2e+00 \\
PAR6 & J0524+4244 & Paloma & 81.1269347122 & 42.7473501585 & 0.11 & RK/Downes & 2MASS J05243042+4244506 & 81.1269347122 & 42.7473501585 & 6.7e-01 \\
PAR7 & J0812+1911 &  & 123.2368971041 & 19.1994030617 & 0.16 & RK/Downes & 2MASS J08125686+1911579 & 123.2368971041 & 19.1994030617 & 8.0e-01 \\
PAR8 & J0813+2813 &  & 123.4667487539 & 28.2214739340 & 0.12 & RK/Downes & 2MASS J08135201+2813174 & 123.4667487539 & 28.2214739340 & 4.0e-01 \\
PAR9 & J1659+1927 &  & 254.9653522369 & 19.4626690472 & 0.14 & RK/Downes & 2MASS J16595166+1927454 & 254.9653522369 & 19.4626690472 & 4.5e-01 \\
PAR10 & PW Aqr &  & 312.5744166667 & -5.6074722222 & 0.07 & RK/Downes & 2MASS J20501785-0536269 & 312.5744166667 & -5.6074722222 & 9.0e-01 \\
PAR11 & V476 Peg & Var 79 Peg & 328.6402235849 & 35.8382128916 & 0.06 & RK/Downes & 2MASS J21543365+3550176 & 328.6402235849 & 35.8382128916 & 8.0e-01 \\
PAR12 & J2337+4308 &  & 354.4967254806 & 43.1475086097 & 0.16 & RK/Downes & 2MASS J23375921+4308509 & 354.4967254806 & 43.1475086097 & 9.8e-01 \\
PAR13 & J1125-0016 &  & 171.4821329930 & -0.2773749807 & 0.06 & RK/Downes & 2QZ J112555.7-001638 & 171.4821329930 & -0.2773749807 & 5.8e-01 \\
PAR14 & J1240-0159 &  & 190.2417827688 & -1.9886640262 & 0.03 & RK/Downes & 2QZ J124058.0-015920 & 190.2417827688 & -1.9886640262 & 4.6e-01 \\
PAR15 & J1304+0103 &  & 196.1740239368 & 1.0585616656 & 0.06 & RK/Downes & 2QZ J130441.7+010330 & 196.1740239368 & 1.0585616656 & 5.4e-01 \\
PAR16 & J1422-0221 &  & 215.7346494973 & -2.3522188534 & 0.14 & RK/Downes & 2QZ J142256.3-022108 & 215.7346494973 & -2.3522188534 & 1.6e+00 \\
PAR17 & J1424-0227 &  & 216.1622066137 & -2.4609022029 & 0.16 & RK/Downes & 2QZ J142438.9-022739 & 216.1622066137 & -2.4609022029 & 1.1e+00 \\
PAR18 & J2224-2924 &  & 336.0677859488 & -29.4060171945 & 0.06 & RK/Downes & 2QZ J222416.2-292421 & 336.0677859488 & -29.4060171945 & 5.3e-01 \\
PAR19 & J2256-2743 &  & 344.1993544274 & -27.7236845659 & 0.35 & RK/Downes & 2QZ J225647.8-274325 & 344.1993544274 & -27.7236845659 & 6.5e-01 \\
PAR20 & J1025-1542 &  & 156.3426554429 & -15.7060177175 & 0.06 & RK/Downes & ASAS J102522-1542.4 & 156.3426554429 & -15.7060177175 & 4.1e+00 \\
\enddata
\tablecomments{This table is published in its entirety in the electronic edition of the {\it Astrophysical Journal}. A sample is shown here.}
\end{deluxetable*}
\end{longrotatetable}

\twocolumngrid
\section{RS Oph} \label{app:RS_Oph}
\begin{figure*}[b]
    \plotone{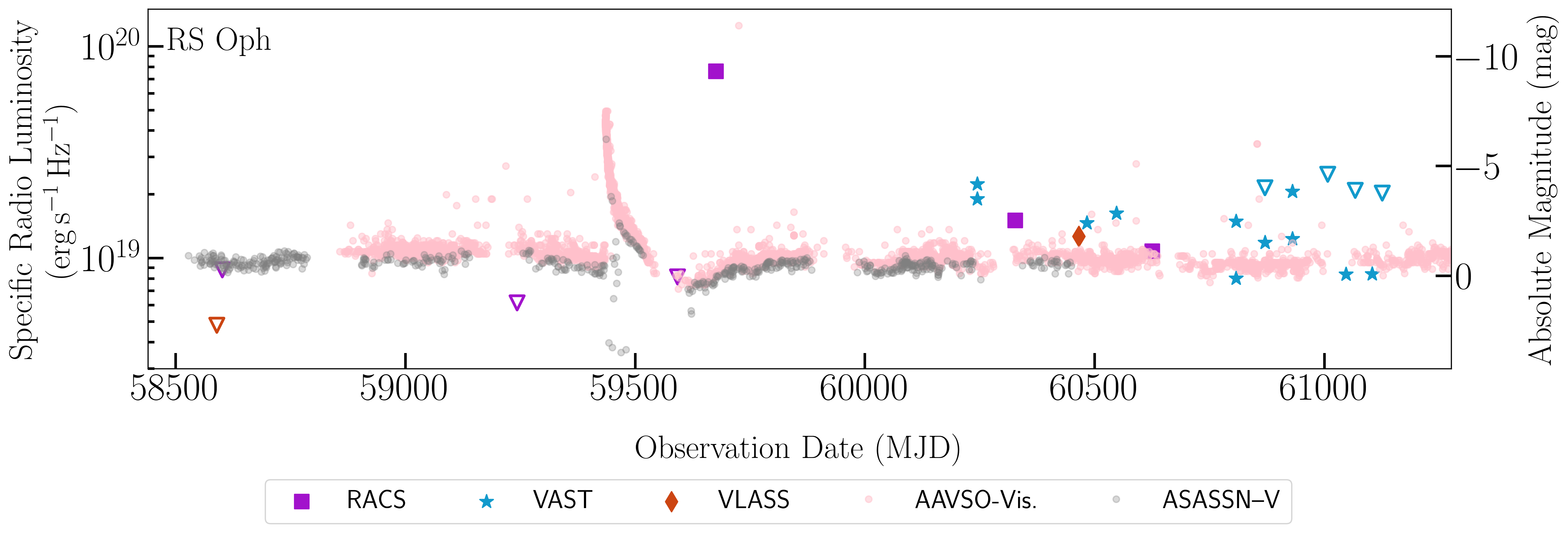}
    \caption{Optical magnitude and specific radio luminosity as a function of time for RS Oph. The optical data plotted in the background consists of AAVSO V-band shown in pink, and ASAS-SN V-band shown in grey. Radio upper-limits in epochs of non-detections are shown with inverted-triangles. RACS, VAST, and VLASS observations are shown as purple squares, blue stars and red diamonds respectively.
    \label{fig:rsoph_lightcurve}}
\end{figure*}

RS Oph, a symbiotic star and one of the most well-studied recurrent novae, goes into a nova outburst every $\approx20\rm\ yr$, including in 2021. It has an orbital period of $\sim453.6\rm \ d$ \citep{Brandi_2009} and lies at a distance of $2.4 \rm \ kpc$. As shown in Figure~\ref{fig:rsoph_lightcurve}, no radio emission was detected from this system prior to the 2021 outburst, whereas radio emission was subsequently detected across all surveys (excluding EMU, which does not have coverage over this region yet). The prominent radio detection immediately following the outburst and the subsequent gradual decline is associated with shock-driven synchrotron emission from the nova ejecta interacting with stellar winds from the red-giant companion \citep[e.g.,][]{Munari_2022, deRuiter_2023, Nayana_2024}.

\begin{deluxetable*}{llllcccccc}[b]
\tabletypesize{\footnotesize}
\tablecaption{Radio detections of RS Oph.
\label{tab:RSOph_radioDetections}}
\tablehead{
\colhead{PARindex} & \colhead{Name} & \colhead{CV type} & \colhead{Survey} & \colhead{Obstime} & \colhead{Radio R.A.} & \colhead{Radio Dec} & \colhead{Sep\tablenotemark{a}} & \colhead{$\rm S_{\nu}$} & \colhead{$\rm L_{R, \nu}$} \\
\colhead{ } & \colhead{ } & \colhead{ } & \colhead{ } & \colhead{ } & \colhead{deg} & \colhead{deg} & \colhead{arcsec} & \colhead{$\rm mJy$} & \colhead{$\rm erg \, s^{-1} \, Hz^{-1}$}
}
\startdata
\multirow{16}{*}{PAR1808} & \multirow{16}{*}{RS Oph} & \multirow{16}{*}{Nova} & VLASS & 2021-11-12T23:08:08.250 & 267.55484 & -6.70795 & 0.02 & 27.09 $\pm$ 0.15 & 1.9 $\pm$ 0.2 $ \times 10^{20}$ \\
& & & RACS & 2022-04-06T20:36:15.200 & 267.55480 & -6.70775 & 0.73 & 11.03 $\pm$ 0.19 & 7.6 $\pm$ 0.8 $ \times 10^{19}$ \\
& & & VAST & 2023-10-27T06:47:48.700 & 267.55514 & -6.70766 & 1.50 & 3.24 $\pm$ 0.58 & $2.2^{0.5}_{0.4}$ $ \times 10^{19}$ \\
& & & VAST & 2023-10-27T07:01:34.900 & 267.55457 & -6.70816 & 1.21 & 2.75 $\pm$ 0.2 & $1.9^{0.3}_{0.2}$ $ \times 10^{19}$ \\
& & & VAST & 2023-10-27T07:01:34.900 & 267.55457 & -6.70816 & 1.21 & 2.75 $\pm$ 0.2 & 1.9 $\pm$ 0.2 $ \times 10^{19}$ \\
& & & RACS & 2024-01-18T02:58:57.400 & 267.55487 & -6.70754 & 1.48 & 2.18 $\pm$ 0.19 & 1.5 $\pm$ 0.2 $ \times 10^{19}$ \\
& & & VLASS & 2024-06-04T09:37:12.450 & 267.55482 & -6.70799 & 0.16 & 1.84 $\pm$ 0.14 & 1.3 $\pm$ 0.2 $ \times 10^{19}$ \\
& & & VAST & 2024-06-21T16:17:38.500 & 267.55673 & -6.70858 & 7.12 & 2.12 $\pm$ 0.71 & 1.5 $\pm$ 0.5 $ \times 10^{19}$ \\
& & & VAST & 2024-06-21T16:17:38.500 & 267.55673 & -6.70858 & 7.12 & 2.12 $\pm$ 0.71 & 1.4 $\pm$ 0.5 $ \times 10^{19}$ \\
& & & VAST & 2024-08-25T11:02:07.000 & 267.55659 & -6.70752 & 6.46 & 2.36 $\pm$ 0.63 & 1.6 $\pm$ 0.5 $ \times 10^{19}$ \\
& & & VAST & 2024-08-25T11:02:07.000 & 267.55659 & -6.70752 & 6.46 & 2.36 $\pm$ 0.63 & $1.6^{0.5}_{0.4}$ $ \times 10^{19}$ \\
& & & RACS & 2024-11-11T07:19:00.700 & 267.55485 & -6.70785 & 0.36 & 1.56 $\pm$ 0.21 & 1.1 $\pm$ 0.2 $ \times 10^{19}$ \\
& & & VAST & 2025-05-12T16:50:21.000 & 267.55451 & -6.70799 & 1.20 & 2.16 $\pm$ 0.62 & $1.5^{0.5}_{0.4}$ $ \times 10^{19}$ \\
& & & VAST & 2025-05-12T16:50:21.000 & 267.55451 & -6.70799 & 1.20 & 2.16 $\pm$ 0.62 & 1.5 $\pm$ 0.5 $ \times 10^{19}$ \\
& & & VAST & 2025-05-12T17:30:29.700 & 267.55474 & -6.70713 & 2.98 & 1.16 $\pm$ 0.24 & 8 $\pm$ 2 $ \times 10^{18}$ \\
& & & VAST & 2025-07-14T14:29:26.700 & 267.55489 & -6.70792 & 0.22 & 1.71 $\pm$ 0.25 & 1.2 $\pm$ 0.2 $ \times 10^{19}$ \\
& & & VAST & 2025-09-12T09:00:15.500 & 267.55479 & -6.70710 & 3.07 & 2.99 $\pm$ 0.64 & 2.0 $\pm$ 0.5 $ \times 10^{19}$ \\
& & & VAST & 2025-09-12T09:13:11.900 & 267.55496 & -6.70810 & 0.67 & 1.79 $\pm$ 0.24 & 1.2 $\pm$ 0.2 $ \times 10^{19}$ \\
& & & VAST & 2026-01-07T01:37:35.800 & 267.55557 & -6.70810 & 2.67 & 1.21 $\pm$ 0.29 & 8 $\pm$ 2 $ \times 10^{18}$ \\
& & & VAST & 2026-03-04T23:33:28.000 & 267.55487 & -6.70801 & 0.24 & 1.22 $\pm$ 0.22 & 8 $\pm$ 2 $ \times 10^{18}$ \\
\enddata
\tablenotetext{a}{Angular separation between the proper-motion-corrected Gaia position and the radio centroid position at that observation epoch.}
\end{deluxetable*}

\end{document}